\documentclass[12pt,preprint]{aastex}

\lefthead{Yorke \& Sonnhalter}
\righthead{Formation of Massive Stars}

\begin{document}

\newcommand\sol{$_\odot$}                      

\title{ON THE FORMATION OF MASSIVE STARS}

\author{Harold W. Yorke}
\affil{MS 169-506, Jet Propulsion Laboratory, 
            California Institute of Technology, Pasadena, CA 91109}
  \email{Harold.Yorke@jpl.nasa.gov}

\and

\author{Cordula Sonnhalter}
\affil{Institut f\"ur Theoretische Physik, Lehrstuhl f\"ur Astronomie,
           Am Hubland, D-97074 W\"urzburg, Germany}

\begin{abstract}

We calculate numerically the collapse of slowly
rotating, non-magnetic, massive molecular clumps of masses 30~M\sol,
60~M\sol, and 120~M\sol, which conceivably
could lead to the formation of massive stars.
Because radiative acceleration on dust grains plays a critical
role in the clump's dynamical evolution, we have
improved the module for continuum radiation transfer in an existing
2D (axial symmetry assumed) radiation hydrodynamic code.  In particular,
rather than using ``grey'' dust opacities and ``grey'' radiation
transfer, we calculate the dust's wavelength-dependent absorption and
emission simultaneously with the radiation density at each wavelength
and the equilibrium temperatures of three grain components: amorphous
carbon particles, silicates, and ``dirty ice''-coated silicates.
Because our simulations cannot spatially resolve the innermost
regions of the molecular clump, however, we
cannot distinguish between the formation of a dense central cluster
or a single massive object.  Furthermore, we cannot exclude significant
mass loss from the central object(s) which may interact with the inflow
into the central grid cell.  Thus, with our basic assumption that all
material in the innermost grid cell accretes onto a single object, 
we are only able to provide an upper limit to the mass of stars which
could possibly be formed.

We introduce a semi-analytical scheme for augmenting existing
evolutionary tracks of pre-main sequence protostars by including the
effects of accretion.  By considering an open outermost boundary, an
arbitrary amount of material could, in principal, be accreted onto this
central star.  However, for the three cases considered
(30~M\sol, 60~M\sol, and 120~M\sol\ 
originally within the computation grid), radiation acceleration limited
the final masses to 31.6~M\sol, 33.6~M\sol, and 42.9~M\sol, respectively,
for wavelength-dependent radiation transfer and to 19.1~M\sol, 20.1~M\sol,
and 22.9~M\sol\ for the corresponding simulations with grey radiation
transfer.
Our calculations demonstrate that massive stars can in principle
be formed via accretion through a disk.  The accretion rate onto
the central source increases rapidly after one initial free-fall
time and decreases monotonically afterwards. By enhancing the
non-isotropic character of the radiation field the
accretion disk reduces the effects of radiative acceleration
in the radial direction --- a process we denote the ``flashlight
effect''.  The flashlight effect is further amplified in our case
by including the effects of frequency dependent radiation transfer.
We conclude with the warning that a careful treatment
of radiation transfer is a mandatory requirement for realistic
simulations of the formation of massive stars.

\end{abstract}

\section{Introduction}
Although massive stars play a critical role in the production of
turbulent energy in the ISM, in the formation and destruction of
molecular clouds, and ultimately in the dynamical and chemo-dynamical
evolution of galaxies, our understanding of the sequence of events
which leads to their formation is still rather limited. Because of
their high luminosities we can expect: a) radiative acceleration will
contribute significantly to the dynamical evolution during the
formation process and b) the thermal evolution time scales of massive
pre-main sequence objects will be extremely short. Thus, we cannot
simply ``scale up'' theories of low mass star formation. Furthermore,
OB stars form in clusters and associations; their mutual interactions
via gravitational torques, powerful winds and ionizing radiation
contribute further to the complexity of the problem.
 
Even though no massive disk has yet been directly observed around
a main sequence massive star, it is likely that such disks are
the natural consequence of the star formation process even in the high
mass case. In their radio recombination maser studies and CO
measurements Martin-Pintado et al. (1994) do find indirect 
evidence for both an ionized stellar wind and a neutral disk
around MWC349. Moreover, several other high luminosity FIR sources
--- suspected embedded young OB stars ---
have powerful bipolar outflows associated with them (e.g., Eiroa
et al. 1994; Shepherd et al. 2000).  Such massive outflows
are probably powered by disk accretion, and, similar to their
low mass counterparts, the flow
energetics appear to scale with the luminosity of the source
(see Cabrit \& Bertout 1992; Shepherd \& Churchwell 1996; Richer
et al. 2000).

The detailed structure and evolutionary history of massive
circumstellar disks has important consequences with regard to the
early evolution of these protostars. Disks provide a reservoir of
material with specific angular momentum too large to be directly
accreted by the central object. Only after angular momentum is
transported outwards can this material contribute to the final mass of
the star. The transition region disk-star will strongly influence the
star's photospheric appearance and how the star interacts with the
disk. The relative high densities in these disks provide the
environment for the further growth and evolution of dust grains,
affecting the disk's opacity and consequently its energetics and
appearance. The disk can be expected to interact with stellar outflows
and is likely to be directly responsible for the outflows associated
with star formation. 

Disks surrounding massive stars or disks associated with close
companions to massive stars should be short-lived compared to their
low mass counterparts.  The UV environment within an OB cluster will
lead to the photoevaporation of disks on a time scale of a few
10$^5$~yr (Hollenbach, Yorke, \& Johnstone 2000).  Because this
process operates on a time scale comparable to the formation of
massive stars and is competitive to it, it is
important to carefully model the transfer of radiation in the
envelopes of accreting massive stars.  Numerical tools capable of
this task are lacking at present.  We consider the present investigation
as an important step in this direction.

\section{The numerical model}\label{numerics}

We consider the simulation of the hydrodynamic collapse
of a rotating molecular cloud clump, with wavelength-dependent
radiation transfer,
under the assumption of symmetry with respect to the rotation axis and
the equatorial plane.  Our code contains all of the basic features of
the 2-D code described in detail by Yorke \& Bodenheimer (1999;
hereafter YB).  In the following we shall discuss only the deviations
from and enhancements to the YB code.

In addition to artificial viscosity for
the treatment of shocks, physical viscosity has been implemented via an
$\alpha$ prescription (Shakura \&\ Sunyaev 1973):
$\nu = \alpha c_S(r) H(r)$, where we have approximated the local disk
scale height by $H(r) = c_S(r)/\Omega (r)$ ($\Omega$ is the angular
velocity and $c_S$ is the sound speed).  All components of 
the viscosity tensor are included, as previously implemented in 2-D 
disk models by R\'o\.zyczka et al. (1994) and YB. Contrary to YB, who
allowed $\alpha$ to vary in time in order to mimic the effects of
tidal torques due to the growth of non-axisymmetric gravitational
modes, the value for $\alpha = 10^{-3}$ is was kept constant in space
and time for all cases considered.

The details of this angular momentum transfer scheme --- within
certain limits --- do not critically affect our results, however,
because our central computational zone (where angular momentum
transfer is presumably very critical) is so large. Test calculations
with $\alpha =0.03$ yielded essentially the same results.
By contrast, for $\alpha \le 10^{-4}$ ring instabilities developed
in the disk which led to a premature ending of the calculations.
As discussed by Yorke, Bodenheimer, Laughlin (1995), who did not
include the effects of angular momentum transfer (i.e., $\alpha =0$),
these ring instabilities are unrealistic.  Such a ring would be 
unstable on a very short (dynamical) time scale and the resulting
clumps would exert tidal torques resulting in angular momentum
transfer.

As in YB we utilize a series of hierarchically nested
grids (Berger \&\ Colella 1989; Yorke \&\ Kaisig 1995).  Whereas
YB considered nesting levels of 6 and $60 \times 60$ grids,
allowing the innermost grid to be $\sim$1/2000 of the cloud radius,
here we have considered nesting levels of 3 only and slightly
larger grids ($64 \times 64$).  We assume constant density
$\rho = \rho_0$ along the outermost radius $r_{\rm max}$ and allow
material to enter into or exit from our computational grid
$R^2 + Z^2 \le r^2_{\rm max}$ ($R$ and $Z$ are the cylindrical
coordinates), based on the sign of the radial component of the
velocity.  By contrast, YB assume a semi-permeable outer
boundary at $r_{\rm max}$: Material with positive radial
velocity can leave the computational grid but no mass was allowed
to enter.

As in YB the boundary of the the innermost cell of our innermost
grid is considered to be semi-permeable: Material can flow
into this cell but cannot flow out of it.  Material entering
this cell is assumed to accrete onto a single central object.
We realize that this ``sink cell'' procedure is a gross
simplification of the physics in the innermost regions of our
computational domain
and a number of possibly important effects are being ignored,
e.g. fragmentation and accretion of material onto multiple
objects and the interaction of the accretion flow with powerful
outflows.  For the cases F30 and G30 (see section \ref{inicond}),
for instance, our innermost cell is a
cylinder of radius 40~AU and height 80~AU, whereas for the cases
F120 and G120 the cell is larger by a factor of four.  Although we
cannot follow the mass flow within and possibly out of this cell,
our simulations do provide upper limits to the amount of
material which is available to be accreted by
the central object: If radiative acceleration prevents the
flow of material into the central sink cell, then
the central object cannot accrete it.

\subsection{Modeling the Central Star}
\label{sec-star}

In contrast to YB we consider a slightly more sophisticated treatment
of the radius $R_*$, luminosity $L_*$, and effective
temperature $T_{\rm eff}$ of the central object.  Its mass $M_*$
is uniquely determined by integrating the mass flux into the center
sink cell:
\begin{equation}\label{intM}
  M_* = \int \dot M_{\rm *}\; dt \; .
\end{equation}
We can express the total energy of the central core in terms of a
``structure parameter'' $\eta$:
\begin{equation}
  E_{\rm tot} = - \eta {G M_*^2 \over R_*} \; .
\end{equation}
In principal, $\eta$, a parameter describing the compactness of the
hydrostatic core, must be calculated by solving for the stellar
structure of the accreting hydrostatic core.  For polytropes of
degree $n$, $\eta$ can be derived analytically
(e.g. Kippenhahn \& Weigert 1990):
\begin{equation}\label{eta}
  \eta = {3 \over 10 - 2n}
\end{equation}
A fully convective pre-main sequence protostar can be approximated by
an $n=3/2$ polytrope and $\eta = 3/7$.  As the star approaches the main
sequence, a greater proportion of it becomes radiative, its core
becomes more compact, and $\eta$ increases.

For purposes of discussion we shall assume for the moment that
$\eta = \eta(M_*,R_*)$ is a known function.
In this case the intrinsic core luminosity $L_*$ (which does not
include the contribution to the total luminosity
emitted in the accretion shock front and dissipated within the disk)
is given by:
\begin{eqnarray}\label{Lcore}
  L_* &=& L_{\rm nuc} - \dot E_{\rm tot}
         - \beta {G M_* \dot M_* \over R_*} \cr
      &=& L_{\rm nuc} - E_{\rm tot} \left[
       {\dot \eta \over \eta}
       + \left( 2 - {\beta \over \eta} \right) {\dot M_* \over M_* }
       - {\dot R_* \over R_* } \right]
\end{eqnarray}
where $L_{\rm nuc}$ is the contribution from nuclear burning.

In the following we will assume that $\beta = 1$:  The material
accreted by the star is being added {\sl ever so gently} at its
current radius $R_*$ and that this material adds negligible
entropy to the star.\footnote{For the newly accreted material we
must subtract the difference of potential energy from infinity to
the stellar surface when considering the star's net change of total
energy.  For $\beta=1$ the energy gained by the star due
to heating from the accretion shock (``backwarming'') or by 
dissipating rotational energy within the star is negligible.}
For the total bolometric luminosity of a spherically accreting
star, however, we must include the
contribution of the potential energy of infalling material as it
is dissipated on its way to the stellar surface:
\begin{equation}\label{Lbol}
  L_{\rm bol} = L_* + L_{\rm acc} = L_* + \beta {G M_* \dot M_* \over R_*}
\end{equation}

Equation \ref{Lcore} shows that for constant $L_* - L_{\rm nuc}$
the star should increase or decrease its radius due to mass
accretion, depending on the sign of the coefficient of $\dot M_* / M_*$.
Because $\beta \approx 1$ and $\eta \approx 3/7$ during the
fully convective Hayashi phase, this coefficient is negative
($\approx -1/3$).  Thus, mass accretion onto a fully convective star
has a tendency to decrease the star's radius, whereas close to the
main sequence, where $\eta$ is closer to unity, mass accretion
has a tendency to cause the star to bloat up (see Kippenhahn \&
Hofmeister 1977 for a more detailed discussion).  In reality,
the internal readjustment of the star after it has gained mass
also affects the nuclear burning rate and thus has an effect on
both radius and luminosity, depending on the magnitude of the mass
accretion rate and the star's current position in the
Hertzsprung-Russell (HR) diagram.

A classical problem of the mathematical theory of stellar structure
is the question of whether
for stars of given fixed parameters, say chemical composition,
mass, and radius, there exists one and only one solution of the
basic structure equations.  There is actually no mathematical
basis for the so-called ``Vogt-Russell'' conjecture of uniqueness
and indeed, multiple solutions for the same set of parameters
have been found numerically in some cases (see discussion by
Kippenhahn \& Weigert 1990).  However, for the rather simple
cases considered here, spherically symmetric, quasi-hydrostatic
homogeneous pre-main sequence and young main sequence stars,
knowledge of the star's mass and age at any given time does
allow us to uniquely fix its position in the HR diagram, from which
we can determine $L_*$, $T_{\rm eff}$, and $L_{\rm nuc}$.

For the pre-main sequence phase we shall account for deuterium
burning only and use the following approximate expression:
\begin{equation}\label{LD}
  L_{\rm nuc} = L_D \approx L_0(M_*) \left[ {\chi_D \over \chi_{D,0}} \right]
    \left[ {R_0(M_*) \over R} \right]^p \; ,
\end{equation}
where $L_0$ and $R_0$ are the equilibrium deuterium burning rate and
equilibrium radius for a star of mass $M_*$ at its
``birthline'',\footnote{There are alternate definitions of the
concept of ``birthline''.  Here we use the word to describe
the equilibrium position of a homogeneous, deuterium-burning,
pre-main sequence star with a cosmic abundance of deuterium.}
$\chi_{D,0}$ is the cosmic mass abundance of deuterium, and $\chi_D$
is the star's net deuterium abundance.  We have assumed
$p = 21$, which --- because the star's central density 
$\rho_c \propto R^{-3}$ --- corresponds to $L_D \propto \rho_c^7$.
This insures that a non-accreting star remains close to its birthline
until a significant fraction of its deuterium is consumed.

Assuming instantaneous mixing during accretion, the deuterium mass
fraction $\chi_D$ can be calculated from the following equation:
\begin{equation}\label{chiD}
  {d \chi_D M_* \over dt} = \chi_{D,0} \dot M_*
                     - \epsilon_D L_D \; ,
\end{equation}
where $\epsilon_D L_D$ is the rate of deuterium consumption due to
deuterium burning
($\epsilon_D = 1.76 \times 10^{-19}\, {\rm s^2}\, cm^{-2}$ is a constant).

From equations \ref{Lcore} and \ref{LD} we can derive an expression
for $\dot R_*$:
\begin{eqnarray}\label{dRcore-dt}
  {\dot R_* \over R_*} \hspace{-2mm} &=& \hspace{-2mm}
    \left[ 1 - \eta_R \right]^{-1} \times   \cr
   \hspace{-2mm} && \hspace{-2mm} 
    \left(\left[ 2 - {\beta \over \eta} + \eta_M \right] {\dot M_* \over M_*}
    + {L_* -L_{\rm nuc} \over E_{\rm tot}} \right)
\end{eqnarray}
where
\begin{equation}\label{eta-defs}
    \eta_R = \left({\partial \ln \eta \over \partial \ln R}
    \right)_{\hspace{-1mm} M}
      \quad {\rm and} \quad
    \eta_M = \left({\partial \ln \eta \over \partial \ln M}
    \right)_{\hspace{-1mm} R}
\end{equation}
From knowledge of $\eta(M_*,R_*)$, $L_*(M_*,R_*)$, $L_0(M_*)$, and
$R_0(M_*)$ we approximate the pre-main sequence evolution of an accreting
protostar by integrating equations \ref{chiD} and \ref{dRcore-dt}
simultaneously.

%

\begin{figure}[htbp]
\epsscale{1.0}
\plotone{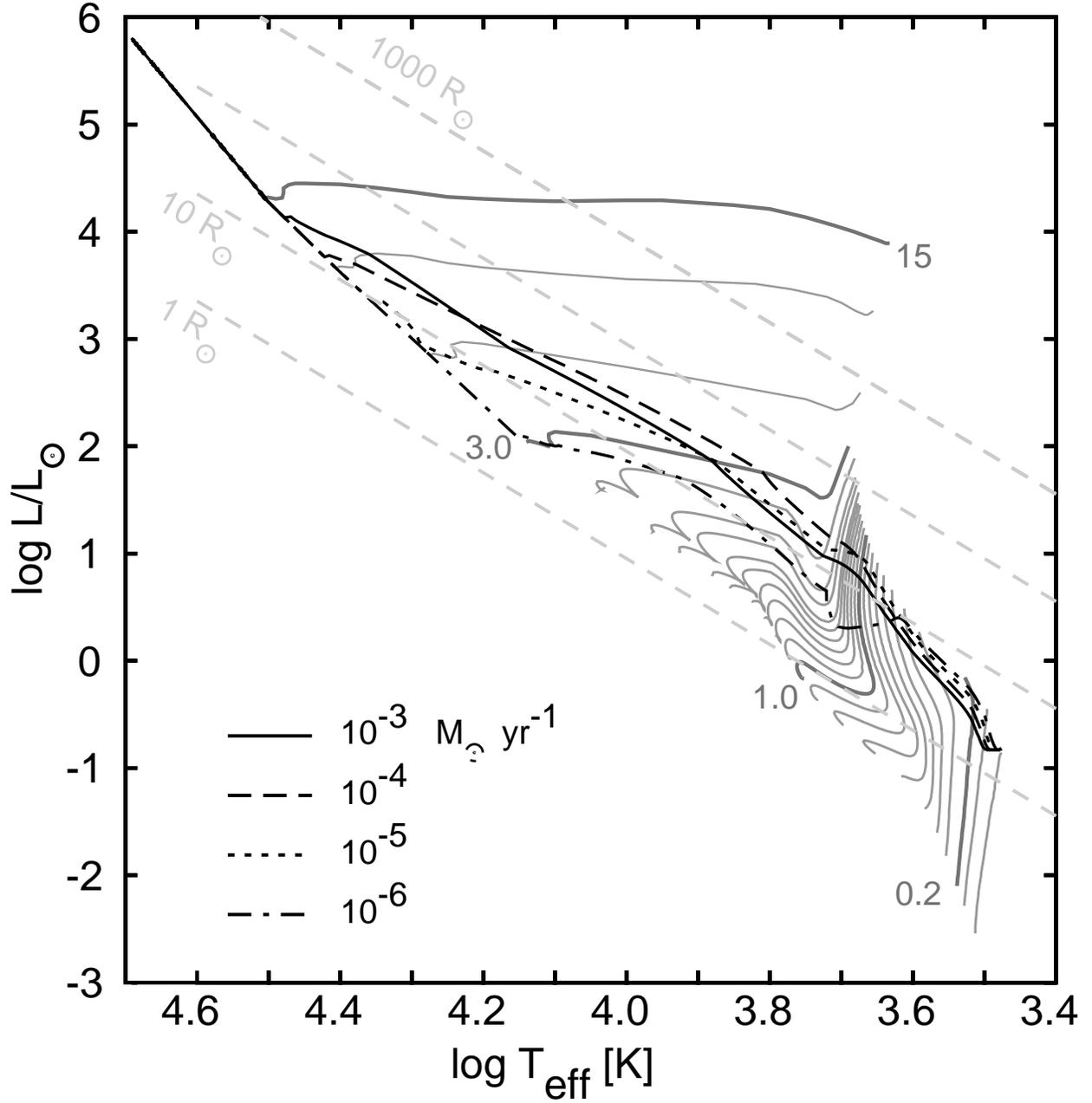}                                        
\caption[]{Evolutionary tracks of pre-main sequence
stars accreting material at a constant rate as indicated.
All accreting tracks begin at $M_* = 0.1$~M\sol\ and 
$R_* = 1$~R\sol.  The total bolometric luminosity, which includes
contributions $L_{\rm acc}$ from the relaxation zone behind the
accretion shock and emission from an accretion disk, is not shown.
For comparison, the evolutionary tracks of non-accreting
stars according to D'Antona \& Mazzitelli (1994) and Iben (1965)
are given ({\it solid grey lines}); the tracks for 0.2~M\sol, 1.0~M\sol,
3.0~M\sol, and 15~M\sol\ are drawn slightly thicker. \\
}
\label{fig-HRD}
\end{figure}

How does one actually determine $\eta$ and $L_*$ from knowledge of
$M_*$ and $R_*$?  For the Hayashi phase we have used published
pre-main sequence tracks\footnote{We use the evolutionary time
$\tau$ given for the published tracks as a parameterization of the
curves.  Our evolutionary time $t$ results from integrating 
equation \ref{dRcore-dt}.} for $L_*(M_*,\tau)$, $R_*(M_*,\tau)$ and set
$\eta = 3/7$.  For the mass range 0.1~M\sol~$\le M_*\le 2.5$~M\sol\ 
we use the evolutionary tracks of D'Antona \& Mazzitelli (1994), which
assume ``CM convection'' ($\alpha_{\rm ML} = 2$) and Alexander +
RI opacities ($Y=0.28$, $Z=0.019$).
For 3~M\sol~$\le M_*\le 15$~M\sol\ we use tracks published by
Iben (1965).  Both sets of tracks represent a series of stellar models
which incorporate the detailed microphysics of convection and stellar
atmospheres.  For masses $M_* > 15$~M\sol\ we have assumed that the
evolutionary tracks for non-accreting stars are horizontal lines at
the main sequence luminosity, similar to the 15~M\sol\ track (see Fig.
\ref{fig-HRD}), and the main sequence values of $R_*(M_*)$ and $L_*(M_*)$
were taken from Allen (1973).


As the contracting protostellar becomes radiative, $\eta$
increases.  Once the star reaches the main sequence hydrogen burning
commences and $\eta \approx$~const.  We approximate $\eta$ during
this phase by approximating the main sequence by polytropic models
with the restriction that the main sequence value $\eta_{MS} \le 1$.
Obviously, this is a very rough
approximation, but because we expect $\dot \eta \approx 0$
and $L_* \approx L_{\rm nuc}$ on the main sequence, we are not 
making a significant error.  Our greatest error arises during the
transition from $\eta = 3/7$ (fully convective) to $\eta_{MS}$, where
we have used interpolated values.

Of course, we could obtain a much better approximation to these physical
parameters by solving the full set of radiation hydrodynamic equations
for an accreting hydrostatic object.  This, however, goes far beyond the
scope of the present investigation.  We merely wish to obtain more realistic
approximations for $L_*$ and $R_*$ than those used by YB which were based
on the pre-main sequence evolution of non-accreting low mass protostars.
Because massive stars require at least one phase of high accretion rates,
typically $\dot M_* \ga 10^{-4}$~M\sol~yr$^{-1}$, the effects of accretion
on the evolution of $L_*$ and $R_*$ cannot be completely neglected.
To illustrate this
quantitatively we display the evolution of (proto-)stars accreting at
given constant rates in the HR diagram (Fig. \ref{fig-HRD}).

These tracks compare very well
with published, more detailed calculations by Behrend \& Maeder (2001)
and by Meynet \& Maeder (2000).  Not only do the tracks lie slightly
below the equilibrium deuterium burning ``birthline'' in all cases, but
the qualitative effect of rapid accretion --- namely to shift the 
tracks to even smaller radii below the ``birthline'' --- occurs both
in our simplified model and the above cited calculations.  The reason
for this is the negative sign of $2 - \beta/\eta$ for fully convective
stars.  The tracks of our simplified model converge to the main sequence 
in a manner similar to those of the detailed calculations.

There are some differences, however.  Behrend \& Maeder and Meynet \&
Maeder {\sl begin} their tracks assuming a fully convective 0.7 M\sol\
star taken to be $7 \times 10^5$~yr old, whereas we show here
that the equilibrium deuterium-burning position at cosmic deuterium
abundance is never reached via accretion for masses $M \la 1$~M\sol.
If you add mass too quickly, the star's radius is reduced as discussed
above.  If you add it too slowly (see $10^{-5}$~M\sol~yr$^{-1}$ track
in Fig. \ref{fig-HRD}), a significant amount of the deuterium is
consumed even before 0.7~M\sol\ has been accreted.  Other differences
are accountable by the different time dependent accretion rates.

We remind the reader that these tracks in the HR diagram do not
reflect the actual observable bolometric luminosities of accreting
protostars.  Much of the accretion luminosity will be indistinguishable
from the intrinsic luminosity of the star.  For our hydrodynamic
simulations we will include the effects of the accretion luminosity
when discussing the star's evolution within the HR diagram.

As in YB we add the accretion luminosity $L_{\rm acc}$ (Adams \& Shu 1986)
to the core's intrinsic luminosity to obtain the total luminosity
\begin{equation}\label{Ltot}
  L_{\rm tot} = L_* + \frac 3 4 {G M_* \dot M_* \over R_*}
\end{equation}
Equation \ref{Ltot} differs from equation \ref{Lbol} (for $\beta=1$),
because approximately 1/4 of the total potential energy of the
accreted material is dissipated within the disk and is already
accounted for by our treatment of viscosity.

Finally, knowing  $L_{\rm tot}$ and $R_*$ allows us to determine
$T_{\rm eff}$ from:
\begin{equation}
  L_{\rm tot} = 4\pi \sigma_{\rm SB} R_*^2 T_{\rm eff}^4 \; .
\label{equ-LTR}
\end{equation}
$\sigma_{\rm SB}$ is the Stefan-Boltzmann radiation constant.
When discussing the evolution of the central star in the HR diagram,
we use $L_*$ rather than $L_{\rm tot}$ in equation \ref{equ-LTR}.

\subsection{The Opacity Model}

\begin{figure}[htbp]
\epsscale{1.0}
\plotone{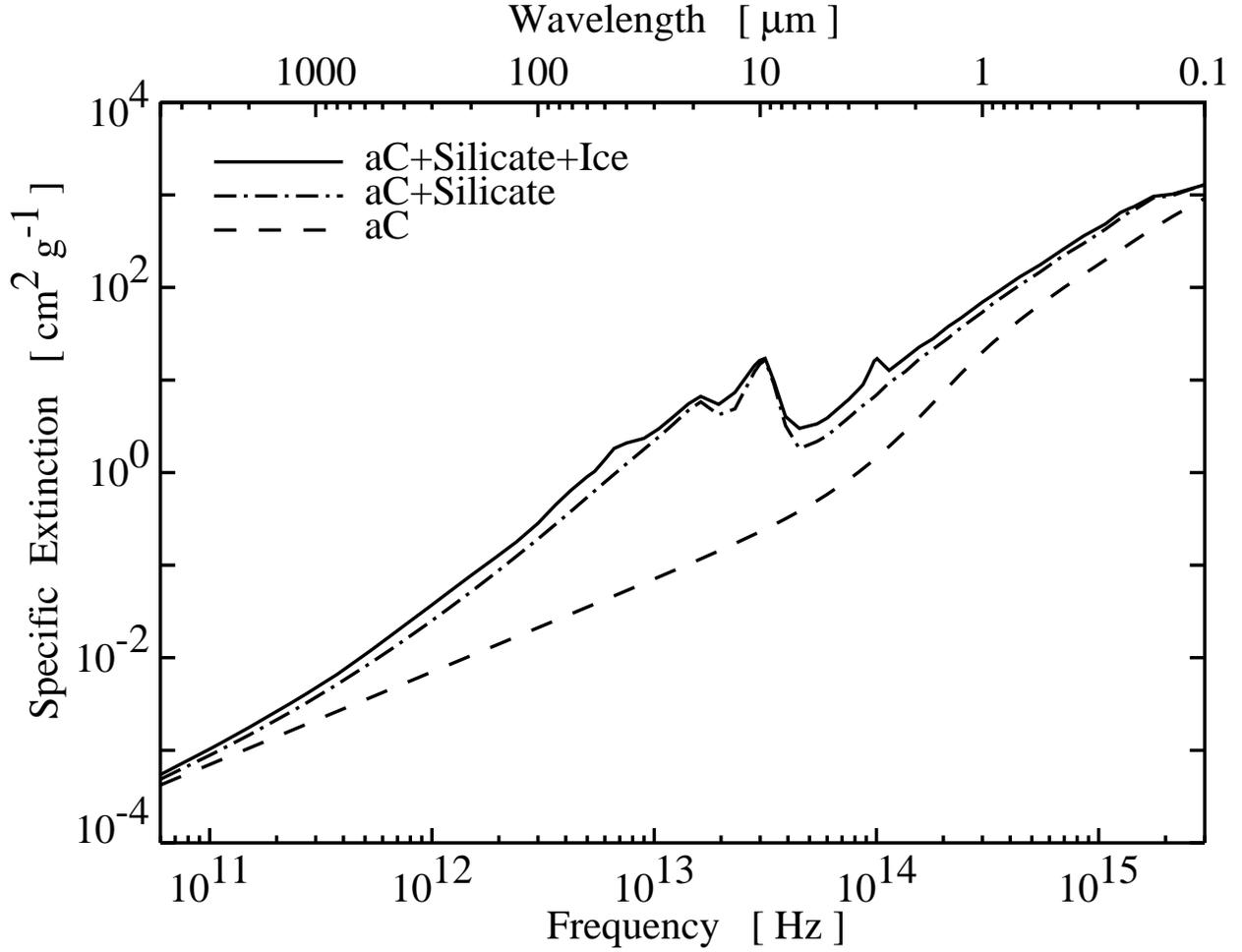}                                         
\caption[]{Specific extinction $\sigma_\nu^{\rm ext}$ of the dust
at temperatures below 125 K ({\it aC + Silicate + Ice}), between
125 K and 1500 K ({\it aC + Silicate}), and between 1500 K and 2000 K
({\it aC}).
}
\label{fig-opacity}
\end{figure}

The principal source of opacity is due to absorption and scattering
by dust (cf. Yorke \& Henning 1994).  We have adopted the detailed
frequency dependent grain model of Preibisch et al. (1993), which
assumes a mixture of small amorphous carbon particles (for grain
temperatures $T_{\rm aC} \le 2000$~K) and ``astrophysical silicate''
grains (for temperatures $T_{\rm Si} \le 1500$~K; c.f. Draine \&
Lee 1984).  At grain temperatures $T_{\rm SiI} < 125$~K
the silicates are coated with a layer of ``dirty'' NH$_3$/H$_2$O ice,
contaminated with 10\% of the amorphous carbon particles.  The grain
sizes are assumed to follow an MRN power law $n(a) \sim a^{-3.5}$
(Mathis, Rumpl, \& Nordsieck, 1977)
in the size ranges 7nm~$\le a_{\rm aC} \le 30$nm (carbon particles)
and 40nm~$\le a_{\rm Si} \le 1\mu$m (silicates).  As evident in Figure
\ref{fig-opacity}, the specific extinction is strongly wavelength
dependent.

For all our calculations we used 64 non-uniformly distributed
frequency points between 5000\, $\mu$m and 0.1\, $\mu$m.

\subsection{Modeling Radiation Transfer}

A complete description of the radiation field requires knowledge
of the radiation intensity $I_\nu({\bf x},\mbox{\bf \^n},t)$,
where {\bf x} indicates the position and {\bf \^n}
the direction under consideration.  Even assuming an axially
symmetric configuration and plane symmetry with respect to the
equatorial plane, $I_\nu$ is a function of 6 independent variables:
frequency $\nu$, two spatial variables $(R,Z)$, two direction
variables, say $(\theta,\phi)$, and time $t$.  Solving the rather
innocuous looking equation for radiation transfer
\begin{equation}
 \frac 1 c \frac{d I\nu} {dt} + \nabla \cdot I_\nu
   = \kappa_\nu^{\rm ext}\, (S_\nu - I_\nu)
\label{radtrans}
\end{equation}
($S_\nu$ is the source function and $\kappa_\nu^{\rm ext} = \sum_i
\kappa_{\nu,i}^{\rm ext}$ is the net contribution to the extinction
coefficient from all components ``i'') together with the equations for
hydrodynamics, energy balance, radiation equilibrium, and the
Poisson equation for the
gravitational potential becomes an formidable task with
present day computers due to the vast number of computations
which are necessary to obtain reasonable numerical resolution.

\subsubsection{Flux-limited diffusion}

The numerical problem can be greatly simplified by resorting to
the flux-limited diffusion (FLD) approximation (see Yorke \&\ Kaisig
1995, Levermore \&\ Pomraning 1981).
If we denote by $J_\nu$ and ${\bf H}_\nu$
the zeroth and first moments of the radiation field, respectively, where
\begin{equation}\label{moments}
    J_\nu = \frac 1 {4\pi} \int I_\nu \, d\Omega \quad {\rm and} \quad
    {\bf H}_\nu = \frac 1 {4\pi} \int I_\nu \, \mbox{\bf \^n} \, d\Omega \; ,
\end{equation}
then the time independent form of the zeroth moment equation --- obtained
by integrating equation \ref{radtrans} over all directions --- can be
written:
\begin{equation}\label{0th-moment}
    \nabla \cdot {\bf H}_\nu
     = j_\nu + \sum_i \kappa_{\nu,i}^{\rm abs}\, [B_\nu(T_i) - J_\nu] \; .
\end{equation}
$\kappa_{\nu,i}^{\rm abs}$ is the absorption coefficient of grain
type ``i'', $T_i$ is its temperature, $B_\nu(T)$ is the Planck function,
$j_\nu$ is the contribution to the emissivity from internal sources
other than dust emission.  The next higher order moment equation,
obtained by multiplying the time independent form of
equation \ref{radtrans} by {\bf \^n} and then integrating 
over all directions, would contain terms involving the second moment
of the radiation field.

The FLD approximation is a procedure for
closing this series of moment equations by approximating the second
moment in terms of the zeroth and first
moments, utilizing knowledge of the opacity distribution:
\begin{equation}\label{H-FLD}
   {\bf H}_\nu = - \frac{\lambda_\nu}{\omega_\nu \kappa_\nu^{\rm ext}}
    \nabla J_\nu \; .
\end{equation}
The effective albedo $\omega_\nu$ and the flux limiter $\lambda_\nu$
are defined as ($\kappa_{\nu,i}^{\rm sca}$ is the scattering coefficient
for grain type ``i''):
\begin{equation}\label{FLD-omega}
  \omega_\nu =\sum_i \frac{\kappa_{\nu,i}^{\rm abs} B_\nu(T_i)
   + \kappa_{\nu,i}^{\rm sca} J_\nu}{\kappa_{\nu,i}^{\rm ext} J_\nu}
\end{equation}
\begin{equation}\label{FLD-dlamda}
  \lambda_\nu = \frac{1}{d}\left(\coth d - \frac{1}{d} \right)
 \quad {\rm with} \quad
  d = \frac{|\nabla J_\nu|}{\kappa_\nu^{\rm ext} \omega_\nu J_\nu}
\end{equation}
In the limits of high opacity or low opacity, the FLD approximation
asymptotically approaches the diffusion limit or the free-streaming
limit, respectively, as expected.

The equations \ref{0th-moment} and \ref{H-FLD} have to be solved for
all frequencies simultaneously with the condition for radiative
equilibrium of each of the absorbing species:
\begin{equation}\label{rad-equilibrium}
 \int \kappa_{\nu,i}^{\rm abs}
  \left[ B_\nu (T_i) - J_\nu \right]\, d\nu = 0 \; .
\end{equation}
When considering ``grey'' radiation transfer as YB did, this
condition can be put into tabular form (for rapid look-up):
\begin{equation}\label{rad-equil}
 T_i = {\cal T}_i(J) \quad {\rm with} \quad J=\int J_\nu\; d\nu \; .
\end{equation}

\subsubsection{Finite Difference Equations}

Following a ``staggered mesh'' discretization philosophy for converting
partial derivatives into finite differences, we define $J_\nu$ at the cell
centers of an $(R_j,Z_k)$ cylindrical grid and the vector
${\bf H_\nu} = (H_{\nu,R},H_{\nu,Z})$ at the centers of cell boundaries.
Combining the equations \ref{0th-moment} and \ref{H-FLD} we obtain
a diffusion equation
\begin{equation}\label{FLD}
  \nabla \cdot \left({\cal D}_\nu \nabla J_\nu \right) = j_\nu +
  \sum_i \kappa_{\nu,i}^{\rm abs} \left[ J_\nu - B_\nu(T_i) \right] \; .
\end{equation}
The discretized form of this FLD equation can be expressed
as a matrix equation:
\begin{equation}\label{FLD-matrix}
 {\cal A}_\nu {\bf J}_\nu = \epsilon_\nu({\bf T}) \; ,
\end{equation}
where the vector ${\bf J}_\nu$ represents the solution for the zeroth
moment and $\epsilon_\nu({\bf T})$ the dust temperature dependent
components for the source terms at all grid centers.
We employ an equidistant $(R_j,Z_k)$ grid at each level of nesting and,
while utilizing a 5-point discretization scheme,
\begin{eqnarray}\label{FLD-5point}
 &a_{l,j,k}^{01} J_{l,j-1,k} + a_{l,j,k}^{10} J_{l,j,k-1} +
  a_{l,j,k}^{11} J_{l,j,k}                                    \cr
 &+ a_{l,j,k}^{21} J_{l,j+1,k} + a_{l,j,k}^{12} J_{l,j,k+1}   \cr
 &= j_{l,j,k}-\sum_i \kappa_{l,i,j,k}^{\rm abs} B_l(T_{i,j,k}) \; ,
\end{eqnarray}
insure by proper centering that our finite difference equation
\ref{FLD-matrix} is accurate to second order ${\cal O}(h^2)$ of
the grid spacing $h = R_{j+1}-R_j = Z_{k+1}-Z_k$.  In equation
\ref{FLD-5point} we have used the subscript $l$ for frequency
and $i$ for grain type.

Note that ${\cal A}_\nu$ is an implicit function of ${\bf J}_\nu$.
Each element of ${\cal A}_\nu$ for frequency $\nu$ at a grid cell
$(j,k)$ contains, with proper centering, $J_{\nu'}$ dependencies for
all frequencies $\nu'$ and extending
beyond the 5-point discretization star of grid cells $(j,k)$ and
$(j\pm1,k\pm1)$.  As evidenced by equation \ref{rad-equilibrium}
the right hand side of equation \ref{FLD-5point} is also an implicit
function of ${\bf J}_\nu$.
The fact that we must find a new solution iteratively on each
nested grid for each hydrodynamic time step places strong demands
on our solution algorithm.

\subsubsection{Boundary conditions}

Boundary conditions are required for each level of nested grids.
For each grid cell $(R_j,Z_k)$ on the outermost grid which satisfies
the condition $R_j^2+Z_k^2 \ge r^2_{\rm max}$
we specify $J_\nu = w \, B_\nu (T_{\rm out})$, where $T_{\rm out}$
is the color temperature of an external isotropic radiation field
and $w$ is the radiation dilution factor.  For all cases considered
here we choose $w=1$ and $T_{\rm out} = 20\,$K.  Along the outer edges
of interior (fine) grids we use the interpolated value of $J_\nu$ 
from the next (coarser) grid level.
Additional boundary conditions along the rotation axis and equatorial
plane result from the assumed symmetry.

Within the innermost central grid cell of each level of nested grids
we treat the central star as an additional internal source of emissivity
and define $j_\nu$ (see equation \ref{0th-moment}) accordingly:
\begin{equation}\label{eq-bcFs} 
  j_\nu = {\pi R_*^2 \over 2\pi R_1^2 Z_1} B_\nu(T_{\rm eff})~,
\end{equation}
where $R_1$ and $Z_1$ are the radial extent and height of the central
cell.

\subsection{Solution algorithms}
Because of the necessity of solving equations \ref{rad-equilibrium}
and \ref{FLD} repeatedly on several grids during the course of
hydrodynamic evolution, it was imperative to make this module
fast and robust. Several promising algorithms which rely on fine-tuning
adjustable parameters had to be abandoned, because a wide range of
problem classes (optically thin, optically thick, strong density
gradients, emission-dominated, scattering-dominated, etc.) occurred,
which were not well suited to a single set of parameters.
We were able to make vast improvements with
respect to the iterative procedures used by Sonnhalter et al. (1995),
who solved these same frequency dependent equations for
a few selected density configurations.  We sometimes sacrificed
robustness (but never accuracy) for speed but automatically fell
back to slower and more robust iterative schemes when the ``faster''
algorithms failed to converge.  We feel it is useful to discuss our
``failures'' as well as our ``successes'' in our endeavor to 
improve the speed of our frequency dependent radiation transfer module.

\subsubsection{Temperature determination}
In analogy to the transfer of line radiation in stellar atmospheres
we consider the method of approximate $\Lambda$ operators (see e.g.
Cannon 1973a,b and Scharmer 1981).  Rewriting equation \ref{FLD-matrix}
with the operator $\Lambda = {\cal A}^{-1}$, we find the formal solution:
\begin{equation}\label{FS}
  {\bf J}_\nu = \Lambda_\nu \epsilon_\nu ({\bf T})
\end{equation}
A simple $\Lambda$-iteration would entail calculating ${\bf T}^{\rm old}$
and thus $\epsilon_\nu({\bf T}^{\rm old})$ from ${\bf J}_\nu^{\rm old}$ using 
equation \ref{rad-equilibrium}.  From this, a new, improved estimate for
${\bf J}_\nu$ can be calculated: 
${\bf J}_\nu^{\rm new} = \Lambda_\nu \epsilon_\nu ({\bf T}^{\rm old}$).
Replacing ${\bf J}_\nu^{\rm old}$ by ${\bf J}_\nu^{\rm new}$ and
repeating this procedure several times may or may not (usually not)
quickly converge to the equilibrium values for ${\bf T}$ and ${\bf J}_\nu$
satisfying equation \ref{FS}.  This procedure is labelled ``$\Lambda$''
in Table \ref{Titerations}.

In order to improve convergence we consider an appropriate
approximation $\Lambda_\nu^*$ to our operator $\Lambda_\nu$.
Rather than using ${\bf J}_\nu^{\rm old}$ in equation
\ref{rad-equilibrium} we substitute the expression
\begin{equation}\label{Jnew}
  {\bf J}_\nu^{\rm new} = {\bf J}_\nu^{\rm old} + \Lambda_\nu^*
  \left[ \epsilon_\nu({\bf T}^{\rm new})
  - \epsilon_\nu({\bf T}^{\rm old}) \right]
\end{equation}
to obtain the equilibrium conditions for each grain component ``$i$'':
\begin{eqnarray}\label{Tnew}
 &\int \kappa_{\nu,i}^{\rm abs} \left[ B_\nu (T_i^{\rm new}) -
  \Lambda_\nu^* \epsilon_\nu({\bf T}^{\rm new}) \right]\, d\nu \cr
 &=\int \kappa_{\nu,i}^{\rm abs} \left[ {\bf J}_\nu^{\rm old} -
  \Lambda_\nu^* \epsilon_\nu({\bf T}^{\rm old}) \right] \, d\nu \; ,
\end{eqnarray}
which can be rewritten in the form ${\bf G}({\bf T}^{\rm new}) = 0$.
The solution vector ${\bf T}^{\rm new}$ can be obtained by solving
equation \ref{Tnew} using a multidimensional Newton-Raphson procedure.

Our approximate $\Lambda$ iteration procedure entails alternatively
solving equation \ref{Tnew} for ${\bf T}^{\rm new}$ and equation
\ref{FLD-matrix} for ${\bf J}_\nu^{\rm new}$.

There are many possibilities for $\Lambda_\nu^*$,
and much effort has been invested in order to optimize its
construction.  Generally speaking, the choice of a particular
$\Lambda_\nu^*$ is based on performance during numerical
experimentation and varies from problem to problem.  Following
this heuristic approach we considered several different approximate
operators and compared their convergence properties with the
simple $\Lambda$ iteration.  Our test cases were the first 20
time steps of two single grid collapse calculations with
a) $13 \times 13$, b) $128 \times 128$ grid cells, and
c) the results of time step 2900 of the 30 M\sol\ case discussed
in section \ref{results}.  For these test calculations we
utilized the most efficient procedure for solving equation
\ref{FLD-matrix} for ${\bf J}_\nu^{\rm new}$ (discussed below).
The average number of $\Lambda$ iterations necessary
for each approximate operator and for each test case is given in
Table \ref{Titerations}.

\begin{table}[ht]
\caption{TEMPERATURE ITERATIONS}
 \begin{center}
  \begin{tabular}{c|ccc}
    \tableline\tableline\\[-3mm]
          &a               &b                &c                \\
procedure &13x13  &128x128 &3*64x64 \\[1mm]
    \tableline\\[-3mm]
    $\Lambda$  &10.1       &16.4             &57.8             \\
    core-wing  &10.9       &---              &---              \\
    diagonal   &4.3        &10.1             &$>100$           \\
    acc-diag   &4.35       &---              &---              \\
    $\Lambda^S$&---        &---              &10.2             \\[-4mm]
  \end{tabular}
 \end{center}
\tablenotetext{}{{\sc Note.}--- Average over 20 test runs}
\label{Titerations}
\end{table}

Our approximate ``core-wing'' operator is con\-struc\-ted
in analogy to the ``core-wing'' $\Lambda$ operator of Scharmer (1981)
which takes into
account that lines are often optically thin in the wings and optically
thick in the line's core.  The analogy with a more or less
monotonously decreasing
(or increasing) continuum absorption coefficient is to assume a
threshold value for which the dusty material becomes optically thin.
We conducted several numerical experiments with different threshold
values but found little improvement in comparison to the simple
$\Lambda$ iteration procedure.

Defining $\Lambda_\nu^{\rm diag} = {\it diag}({\cal A})^{-1}$
resulted in improved convergence behavior.  Use of additional
accelerator terms as outlined by Auer (1987), however, did not always
lead to further improvement.  Our best convergence results were
attained with the $\Lambda_\nu^S$ operator constructed as follows.
We first approximate $J_{l,j,k}$ by using the diagonal operator
$\Lambda_\nu^{\rm diag}$ in equation \ref{FS}:
\begin{equation}\label{FS-diag}
  J_{l,j,k}^*({\bf T}) = \frac{j_{l,j,k} - \sum_i \kappa_{l,i,j,k}^{\rm abs}
  B_l(T_{i,j,k})} {a_{j,k}^{11}} \; .
\end{equation} 
This approximation is used for the off-diagonal
terms in equation \ref{FLD-5point} to obtain:
\begin{eqnarray}\label{LambdaS}
  &\hspace{-10mm} J_{l,j,k} = ({a_{l,j,k}^{11}})^{-1} \Bigl[
  j_{l,j,k} - \sum_i \kappa_{l,i,j,k}^{\rm abs} B_l(T_{i,j,k})      \cr
  &- a_{l,j,k}^{01} J_{l,j-1,k}^*({\bf T})    
   -  a_{l,j,k}^{21} J_{l,j+1,k}^*({\bf T})          \cr
  &\quad - a_{l,j,k}^{10} J_{l,j,k-1}^*({\bf T})
   -  a_{l,j,k}^{12} J_{l,j,k+1}^*({\bf T}) \Bigr]  \; .
\end{eqnarray}
Because the right hand side depends only on known quantities and
the temperatures ${\bf T}$ this corresponds to the operator equation
\begin{equation}
  {\bf J}_\nu = \Lambda_\nu^S \epsilon_\nu({\bf T}) \;.
\end{equation}
Use of $\Lambda_\nu^S$ sometimes leads to oscillations during
the iterations which we are able to damp by interjecting simple 
$\Lambda$ iterations.

Occasionally, the Newton-Raphson iteration procedure used for
solving equation \ref{Tnew} did not converge.  For these special
cases we had to abandon our approximate $\Lambda$ iterations
in favor of the following more robust but much more CPU intensive
procedure.  If we denote by $\tilde T_i$ our (inaccurate)
estimate of $T_i$ then a temperature correction $\Delta T_i$ can
be determined from:
\begin{equation}\label{alt-rad-equil}
  \Delta T_i = \frac{\sum_\mu w_\mu \kappa_{\mu,i}^{\rm abs} \left[
    J_\mu - B_\mu(\tilde T_i) \right]}
    {\sum_\mu w_\mu \kappa_{\mu,i}^{\rm abs} dB_\mu/dT|_{T=\tilde T_i}} \; ,
\end{equation}
where we have replaced the frequency integration by a weighted
sum with the weights $w_\mu$.  By substituting
$B_\nu(\tilde T_i) + dB_\nu/dT|_{T=\tilde T_i} \Delta T_i$ for
$B_\nu(T_i)$ on the right hand side of equation \ref{FLD}
and replacing $\Delta T_i$ by the expression given in equation
\ref{alt-rad-equil} we find a modified FLD equation:
\begin{eqnarray}\label{alt-FLD}
  &\nabla \cdot \left( {\cal D}_\nu \nabla J_\nu \right) = j_\nu +
 \sum_i \kappa_{\nu,i}^{\rm abs} \Bigl( J_\nu - B_\nu(\tilde T_i) \cr
  &-\frac{dB_\nu}{dT}\bigr|_{T=\tilde T_i}
    \frac{\sum_\mu w_\mu \kappa_{\mu,i}^{\rm abs}
    \left[ J_\mu - B_\mu(\tilde T_i) \right]}
    {\sum_\mu w_\mu \kappa_{\mu,i}^{\rm abs} dB_\mu/dT|_{T=\tilde T_i}}
    \Bigr) \; ,
\end{eqnarray}
which can be expressed in matrix form as
\begin{equation}\label{alt-FLD-matrix}
  {\cal A}_\nu {\bf J}_\nu + \sum_\mu {\cal C}_{\nu \mu} {\bf J}_\mu
   = \chi_\nu({\bf T}) \; .
\end{equation}
The iteration procedure now consists of alternatively solving equations
\ref{alt-rad-equil} and \ref{alt-FLD-matrix} until convergence.
Because this alternative procedure is more CPU-in\-ten\-sive than the
approximate $\Lambda$ iterations described above, it is only used
when absolutely necessary.

\subsubsection{Solution of the partial differential equations}
The partial differential equations \ref{FLD} or \ref{alt-FLD}
have to be solved repeatedly for each hydrodynamic time step
and for each frequency.  The coefficients ${\cal D}_\nu$,
$\kappa_{\nu,i}^{\rm abs}$ and $\epsilon_\nu$ depend on frequency
and position and vary from time step to time step.  The relative
importance of individual terms can vary strongly from position to
position and with time: i.e. both
$$
\bigl| \nabla \cdot {\cal D}_\nu \nabla J_\nu \bigr| \gg
\Bigl| j_\nu + \sum_i \kappa_{\nu,i}^{\rm abs} [ J_\nu - B_\nu(T_i) ] \Bigr|
$$
and
$$
\bigl| \nabla \cdot {\cal D}_\nu \nabla J_\nu \bigr| \ll
\Bigl| j_\nu + \sum_i \kappa_{\nu,i}^{\rm abs} [ J_\nu - B_\nu(T_i) ] \Bigr|
$$
occur.

In addition to the ``alternating direction implicit'' procedure (ADI)
we considered the multi-grid scheme ``full approximation storage''
(FAS), ``successive over-relaxation'' (SOR) and its special case by
Gauss-Seidel (GS), the method of ``quasi-minimal residues'' (QMR),
and bicgstab(2), an improvement on the ``Bi-CGSTAB'' algorithm
(``Bi-CGSTAB'' combines the advantages of GMRES and Bi-CG).  More
information on the ADI, FAS, SOR, and GS procedures can be found
in Press et al. (1992).  Our variant of the QMR procedure is
described by B\"ucker \& Sauren (1996) and the bicgstab(2) method
is described in full detail by Sleijpen \& Fokkema (1993).

\begin{figure*}[htbp]
\epsscale{1.00}
\begin{center}
  \plotone{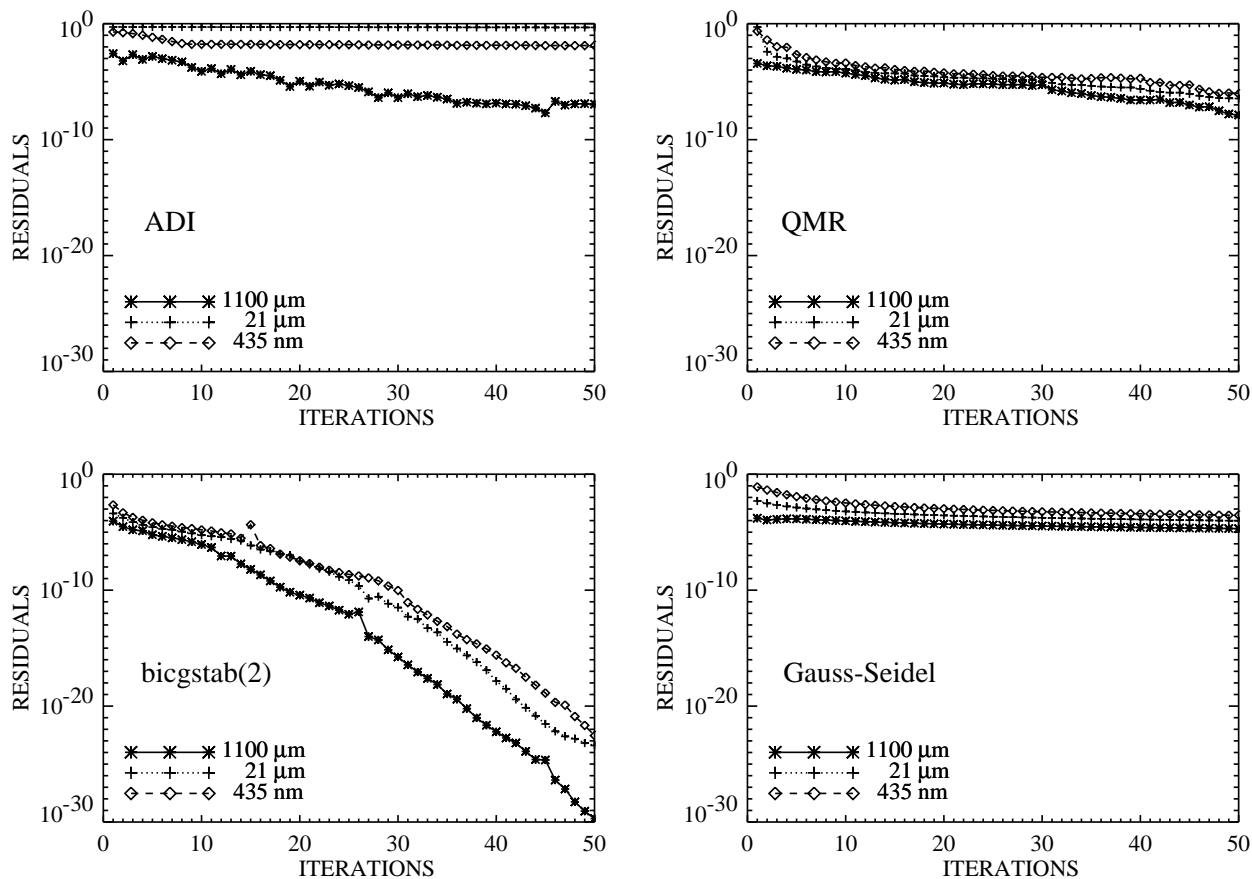}
\end{center}
\caption[]{Convergence properties of four versions of the
partial differential equation solver for equation \ref{FLD}
are shown at three wavelengths: 435\,nm (high optical depth),
21\,$\mu$m (optical depth of order unity), and
1100\,$\mu$m (small optical depth).
The ``residual'' is the Euclidean norm of
$\left|{\cal A}_\nu {\bf J}_\nu - \epsilon_\nu({\bf T}) \right|$
over the entire grid.
}
\label{fig-comparison}
\end{figure*}

Our standard test problem was constructed from the density distribution
calculated by Yorke, Bodenheimer, Laughlin (1995) for the case of a
slowly rotating 10 M\sol\ molecular clump, 7074 years after the
beginning of collapse (see their Fig. 4). The density was remapped 
onto a single $66 \times 66$ grid.  Assuming a central
luminosity of 100~L\sol\ and constant dust temperature ($T_i = 20\,$K)
as starting values, we constructed an initial model by iterating
equation \ref{FLD} until convergence using SOR, subsequently
solving for the corrected temperatures using equation
\ref{rad-equilibrium}.  The time required for this first iterative
step was not included in our comparison.  Each procedure was coded
in fortran90 and optimized for vector processing; a single processor
of a CRAY T90 was used for the comparison calculations, the results
of which are shown in Figure \ref{fig-comparison} and Table
\ref{Tcomparison} for ADI, QMR, bicgstab(2), and GS.

\begin{table}[htb]
\caption{CPU TIME NECESSARY FOR CONVERGENCE}
\begin{center}
  \begin{tabular}{c|cc}
    \tableline\tableline\\[-3mm]
procedure &$1100\,\mu$m    &435\,nm   \\[1mm]
    \tableline\\[-3mm]
    ADI        &2.12       &$>600$    \\
    GS         &3.98       &14.30     \\
    QMR        &2.17       &2.03      \\
    bicgstab(2)&1.15       &1.38      \\[-4mm]
  \end{tabular}
\end{center}
\tablenotetext{}{{\sc Note.}--- Average over 10 tests.
Calculations were run until the residual
fell below $10^{-30}\,$erg s$^{-1}$ cm$^{-3}$ Hz$^{-1}$.}
\label{Tcomparison}
\end{table}

We first note that the ADI procedure used by
Sonnhalter et al. (1995) did not converge for several frequencies for
our test problem; the residual in the central cell remained constant
after a few iterations.
This problem was alleviated by resorting to SOR; the most
robust variant used an over-relaxation parameter $\omega=1$,
reducing SOR to the GS procedure.  GS iterations resulted in 
steadily (but slowly) decreasing residuals.  QMR and bicgstab(2)
required the fewest number of iterations to reach convergence, but
because of their complexity, it is not immediately apparent that
these procedures would also require the least amount of CPU time.
As shown in Table \ref{Tcomparison} bicgstab(2) did indeed prove
to be the most efficient partial differential equation solver.  
It was used for our numerical simulations discussed in section
\ref{results}.  Every variant of FAS we attempted diverged for
our test problem and the method was quickly abandoned.

\section{Initial conditions}\label{inicond}

In spite of recent observational and theoretical progress, the
initial conditions for protostellar collapse are still poorly
known. Whereas it is clear that the formation of massive stars
require high masses, neither the average temperatures nor the
sizes and density distributions of those molecular clumps on the
brink of forming massive stars are especially well known
(Stahler, Palla, \& Ho 2000).

Because we are investigating whether massive stars can form by
accretion, comparable to our current understanding of low mass
star formation rather than by coalescence of lower mass
hydrostatic components or stellar coalescence
(see e.g. Bonnell, Bate, \& Zinnecker 1998),
we will adopt for our initial conditions
a scaled-up version of the initial configuration expected for the
formation of low mass stars (see e.g. Williams, Blitz, \& McKee 2000
or Andr\'e, Ward-Thompson, \& Barsony 2000 for extensive reviews):
clump sizes are a fraction of a pc, temperatures lie in the range
10--30\,K (we adopt $T_0 = 20\,$K),
and the clumps are density-peaked towards the
center (typically, $\rho \propto r^{-p}$, where $p \approx 1-2$).
Because high mass star formation is an extremely rare
event, we do not restrict ourselves to clump masses
which are one or two Jeans masses only.  We adopt a thermal to
gravitational binding energy ratio of $E_T/|E_G| = 0.05$,
corresponding to about 10 Jeans masses.  However, because
gravity always dominated thermal pressure forces in our domain
of integration, we expect no significant evolutionary differences
as long as $T_0 \la 100\,$K ($M \ga 2$~Jeans masses).

\begin{table}[htb]
\caption{INITIAL CONDITIONS FOR COLLAPSE}
\begin{center}
  \begin{tabular}{c|ccc}
    \tableline\tableline\\[-3mm]
\hbox to 35mm{\hfil frequency-dependent}
                     &F30     &F60     &F120      \\
\hbox to 35mm{\hfil grey cases}
                     &G30     &G60     &G120      \\
    \tableline\\[-3mm]
\hbox to 35mm{Mass \hfil [M\sol]}
                     &30      &60      &120       \\
\hbox to 35mm{Radius \hfil [pc]}
                     &0.05    &0.1     &0.2       \\
\hbox to 35mm{$T_0$ \hfil [K]}
                     &20      &20      &20        \\
\hbox to 35mm{$\Omega_0$ \hfil [10$^{-13}$\,s$^{-1}$]}
                     &5       &5       &5         \\
\hbox to 35mm{$\rho_0$ \hfil [$10^{-20}$\,g cm$^{-3}$]}
                     &1       &1       &1         \\
\hbox to 35mm{$\left< n_H \right>$ \hfil [10$^5$\,cm$^{-3}$]}
                     &23      &5.8     &1.5       \\
\hbox to 35mm{$E_T/|E_G|$ \hfil}
                     &0.05    &0.05    &0.05      \\
\hbox to 35mm{$E_{\rm rot}/|E_G|$ \hfil}
                     &0.023   &0.094   &0.374     \\
\hbox to 35mm{$t_{\rm ff}$ \hfil [$10^3$ yr]}
                     &32      &65      &129       \\[-4mm]
  \end{tabular}
\end{center}
\tablenotetext{}{{\sc Note.}--- The initial clump temperature
$T_0$ was also the outer temperature boundary condition
$T_{\rm out}$.}
\label{Tinitial}
\end{table}

The initial configuration is summarized in Table \ref{Tinitial}.
We begin with a rotating ($\Omega = 5\times 10^{-13}\,$s$^{-1}$)
density configuration $\rho \propto r^{-2}$.  This power law
dependence corresponds to that expected in the central regions of
magnetically supported, slowly contracting, slowly-rotating
clouds (Mouschovias 1990; Tomisaka, Ikeuchi, \& Nakamura 1990;
Lizano \& Shu 1989; Crutcher et al. 1994). Probably the best
measured pre-collapse core, L1689B, has a somewhat flatter
central region ($R \la 4000$~AU) with $\rho \propto r^{-0.4}$ or
$\rho \propto r^{-1.2}$, depending on geometric and temperature
assumptions (Andr\'e, Ward-Thompson, \& Motte 1996).  Outside 4000 AU
the density decreases as $\rho \propto r^{-2}$. By contrast, Motte,
Andr\'e, \& Neri (1998) find $\rho \propto r^{-2}$ density profiles
in 10 compact pre-collapse cores in the $\rho$ Oph region, which they
were able to resolve down to $\sim$140~AU.

In spite of the fact that the radial dependence of density in
pre-collapse cores is observationally ill-constrained, especially
for the high mass case, we have restricted ourselves to an initial
$\rho \propto r^{-2}$ density configuration, corresponding
to that of a ``singular isothermal sphere'', which has been used for
many years to model the formation of low mass stars.  This allows
direct comparision with the results of analogous collapse
calculations.
During the course of evolution, however, rotation, thermal forces
at the outermost radius, and radiation forces quickly modify this
initial density distribution.  Moreover, in contrast to similarity
solutions that rely on the assumption of
a singular isothermal sphere our clumps are 
initially at rest.  The mass accretion rates are thus not fixed,
but rather a result of the radiation hydrodyanmic calculations.

For the cases F120 and G120 the rotational velocity at the equator
and $r_{\rm max}$ exceeded the escape velocity.  Thus, about 7.3\,M\sol\
of the original 120\,M\sol\ in the cloud was not bound initially.

\section{Results}\label{results}


\subsection{Evolution of the Central (Proto-)Star}

\begin{figure}[htbp]
\epsscale{0.50}
\plotone{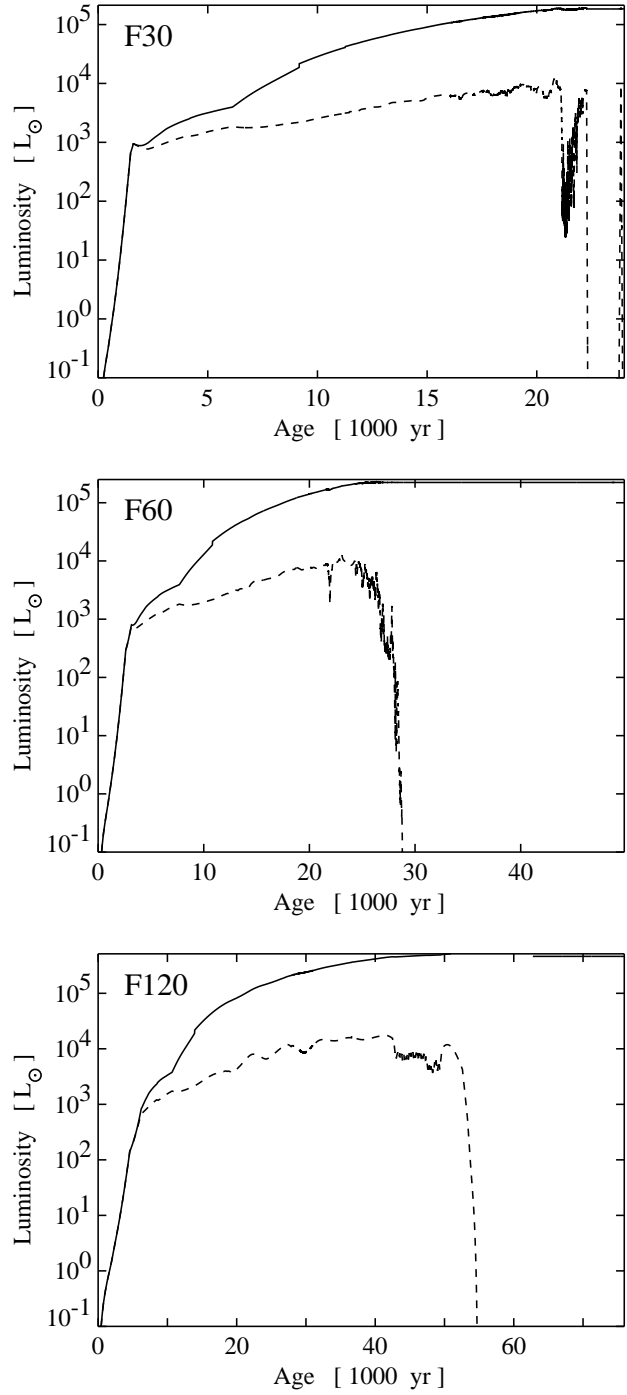}
\caption[]{Total luminosity ({\it solid lines}) and accretion
luminosity ({\it dashed lines}) for the frequency-dependent
``F'' sequences
}
\label{fig-L_evol}
\end{figure}

We have implicitly assumed that all material flowing into the central
zone is accreted onto a single object.  Its mass is known from
integrating the mass accretion rate $\dot M_*$ over time; its
luminosity and effective temperature are modeled according to
the procedure described in section \ref{sec-star}.
In Fig. \ref{fig-L_evol} we display the evolution of the total
and accretion luminosities for the frequency-dependent ``F''
sequences.

\begin{figure}[htbp]
\epsscale{1.00}
\plotone{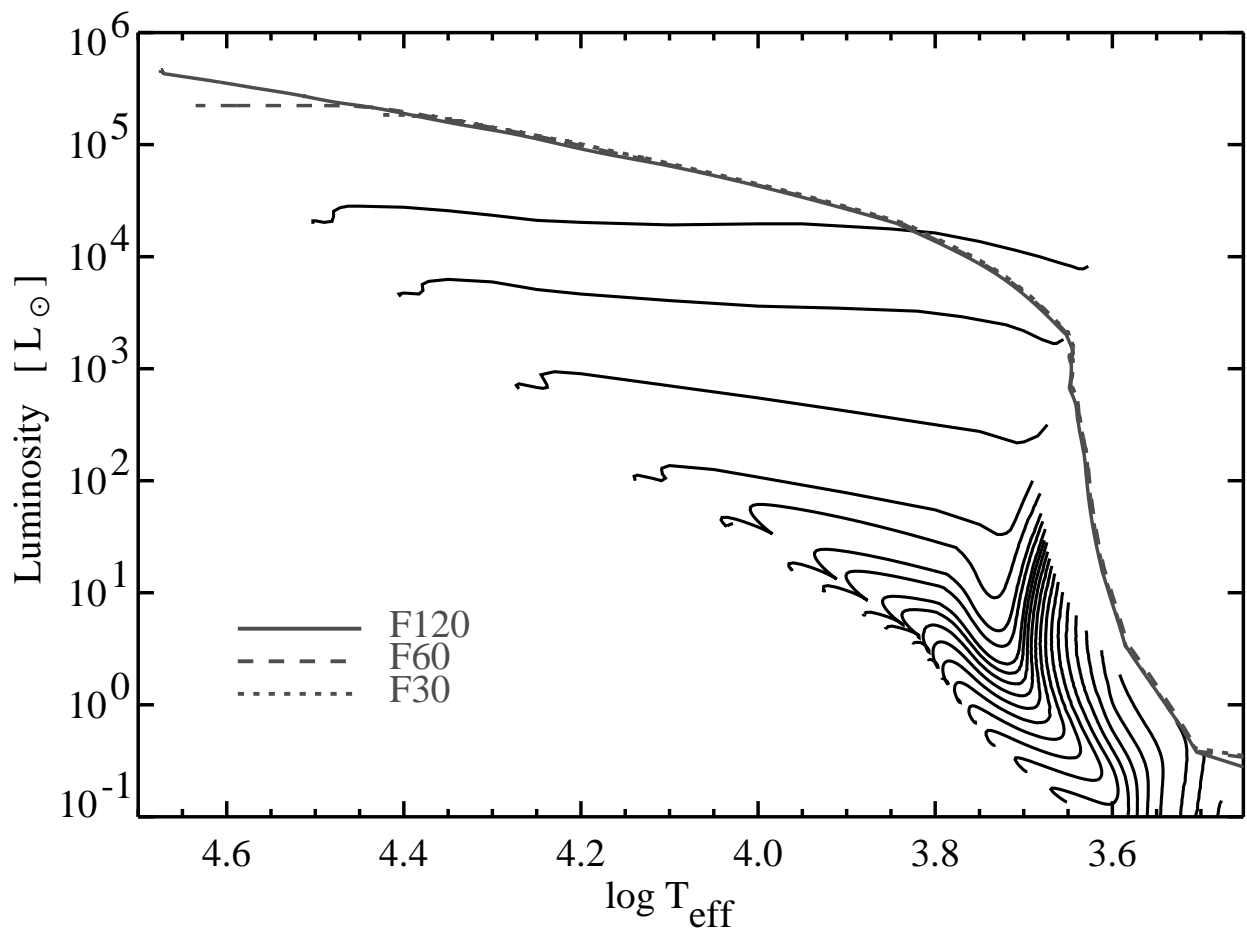}                                         
\caption[]{Evolution of the central (proto-)stars for the
frequency-dependent ``F'' sequences in the HR diagram.  The
luminosity contribution $L_{\rm acc}$ from the relaxation
zone behind the accretion shock has been included.  For
comparison, the evolutionary tracks of non-accreting stars
has been given (c.f. Fig. \ref{fig-HRD}).
}
\label{fig-HRDevol}
\end{figure}

\begin{figure}[htbp]
\epsscale{1.00}
\plotone{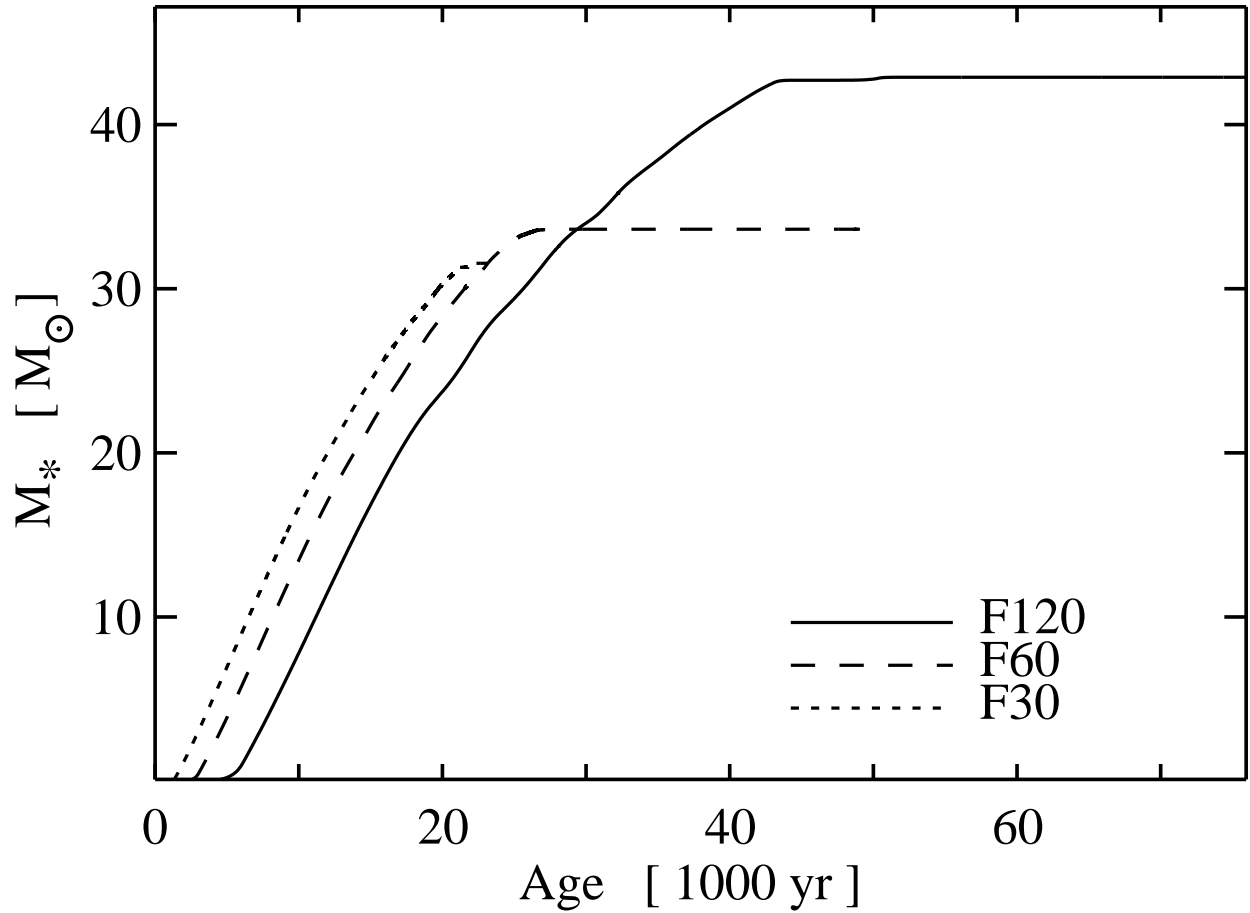}                                         
\caption[]{Evolution of the central (proto-)stars' masses for the
frequency-dependent ``F'' sequences. After an initial delay the
mass accretion rates rise rapidly to comparable values (given by
the slope of the curves) and fall off rapidly as radiative effects
inhibit further accretion (see also Fig. \ref{fig-mass60}).
}
\label{fig-massevol}
\end{figure}

Considering the fact that the initial free-fall times double from
case F30 to F60 and from F60 to F120, the initial rapid rise of
luminosity --- due to the emission from an accretion shock surrounding
the central star --- appears similar for all three cases.  This is
not too surprising, because neither rotation nor radiation has
an important influence during these early phases; the flow is
dominated by gravity.  After several thousand years, the energy
released within the accretion shock front is {\sl not} the
dominant source of luminosity.  Adding material onto the central
star at the very high rates considered here causes it to bloat
up to radii exceeding the deuterium burning limit (see Fig.
\ref{fig-HRDevol}).  Note that there is very little difference
of the evolution within the HR diagram of these
three cases.  Indeed, except for a ``shift'' in time and the
maximum mass attained in each simulation, the growth
of mass within the central computational zone was remarkably
similar (c.f. Fig. \ref{fig-massevol}).  The rate of mass accretion
rises sharply within a few thousand years, reaches a maximum value
$\sim 2 \times 10^{-3}$~M\sol~yr$^{-1}$.

\begin{figure*}[htbp]
\epsscale{1.00}
\plotone{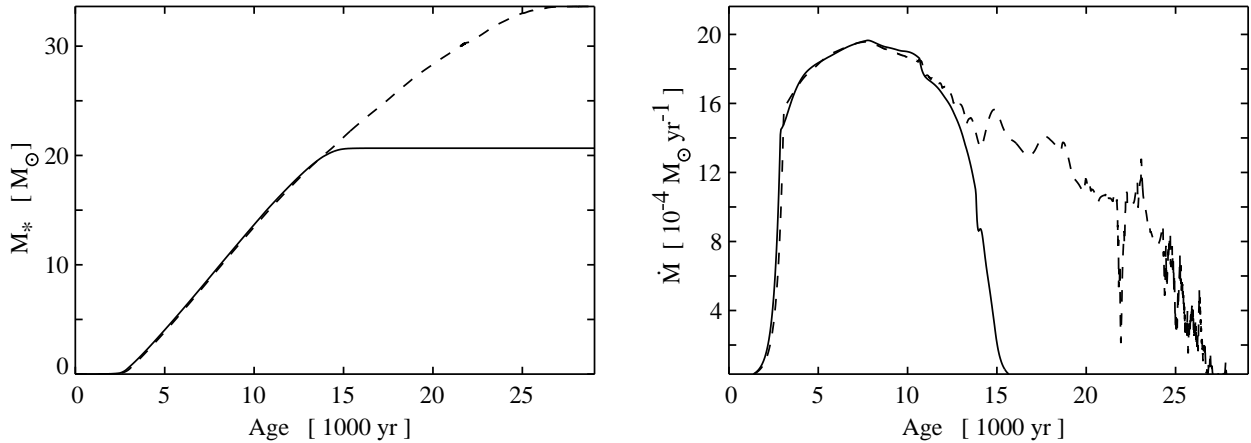}
\caption[]{Evolution of central mass ({\it left frame}) and mass
accretion rate ({\it right frame}) for the frequency-dependent case
F60 ({\it dashed line}; shown previously in Fig. \ref{fig-massevol})
and the grey case G60 ({\it solid line}).
}
\label{fig-mass60}
\end{figure*}

Comparing these tracks with those published by Behrend \& Maeder
(2001) by Meynet \& Maeder (2000), we note that our tracks lie
significantly higher in the HR diagram (larger radii, larger
luminosity, but slightly lower $T_{\rm eff}$ for equivalent
masses) and cross the main sequence at a somewhat higher
luminosity by about 0.5 to 0.8 dex.  This is not too surprising,
because our luminosity includes 3/4 of the accretion luminosity
(see equation \ref{Ltot}) and our accretion rate varied in time
according to the results of hydrodynamic calculations (see
Fig. \ref{fig-massevol} and Fig. \ref{fig-mass60}).  By contrast,
Behrend \& Maeder assumed
a given accretion rate onto the star based on observations of
outflows $\dot M_* = \max(10^{-5}\;{\rm M}_\odot\; {\rm yr}^{-1},
\tilde f\, \dot M_{\rm out})$, where $\tilde f$ was chosen to
lie between 0.3 and 0.5.  For outflow mass loss $\dot M_{\rm out}$
the authors used the observed relation between the outflow mass
rates and the stellar bolometric luminosities in ultracompact
HII regions found by Churchwell (1998) and confirmed by Henning
et al. (2000).  Meynet \& Maeder used the formula
$\dot M_* = 10^{-5}\;{\rm M}_\odot\; {\rm yr}^{-1}
\max(1,M_*/{\rm M}_\odot)^{1.5}$.

The formulae assumed by Behrend \& Maeder and  by Meynet \& Maeder
result in higher accretion times (lower average accretion rates)
and a shift of the phase of extremely high accretion rates to later
evolutionary times, compared to what we find in our collapse
calculations.
At the high mass end, however, the evolution of the star is similar;
the tracks follow closely along the main sequence in spite of high
mass accretion rates.

The growth of mass in the central computational zone was governed
by the relative importance of centrifugal, radiative, and gravitative
forces in the molecular clump.  For the six cases calculated we
found that the detailed treatment of radiation transfer strongly
influenced the clump's evolution.  In general, the accretion rate
increased sharply after about one free-fall time and then decreases.
To exemplify this we display
the growth of central mass and the time dependence of the accretion
rate for the frequency-dependent case F60 and for the grey case
G60 in Fig. \ref{fig-mass60}.  Assuming grey radiation transfer,
in infall of material into the central regions of the molecular
clump is strongly hampered after about 14\,000 years.  The central
star is a 20.7 M\sol\ main sequence star with a luminosity of
$5.2 \times 10^4$~L\sol.  Assuming frequency-dependent radiation
transfer, more than 60\% more mass could accrete onto the central
object. For case F60 the final mass was 33.6~M\sol\ and the final
main sequence luminosity was $2.2 \times 10^5$~L\sol.

\subsection{Evolution of Molecular Cloud Clumps}

\begin{table}[htbp]
\caption{PARAMETERS OF EVOLVING CLUMPS}
\begin{center}
  \begin{tabular}{r|cccc}
    \tableline\tableline\\[-3mm]
case\phantom{XX}&Age       &$M_{grid}$  &$L_*$       &$M_*$       \\
      Fig.      &[1000 yr] &[M$_\odot$] &[1000 L$_\odot$] &[M$_\odot$] \\
                &          &            &            &            \\[-3mm]
    \tableline\\[-3mm]
F60: 8a         &10        &66.2        &11          &13.4        \\
    8b          &20        &73.2        &133         &28.4        \\
    8c          &25        &72.9        &210         &33.0        \\
    8d          &30        &62.2        &223         &33.6        \\
    8e          &35        &47.0        &223         &33.6        \\
    8f          &45        &39.9        &223         &33.6        \\
                &          &            &            &            \\[-3mm]
G60: 9a         &10        &66.4        &14          &13.7        \\
    9b          &25        &69.2        &52          &20.7        \\
    9c          &35        &69.0        &52          &20.7        \\
    9d          &90        &68.8        &52          &20.7        \\
    9e          &100       &68.9        &52          &20.7        \\
    9f          &110       &69.0        &52          &20.7        \\
                &          &            &            &            \\[-3mm]
F30: 10a         &10        &37.9        &27          &16.7        \\
    10b          &15        &41.6        &85          &24.4        \\
    10c          &19        &43.7        &143         &29.0        \\
    10d          &21        &44.6        &178         &31.2        \\
    10e          &22        &44.5        &181         &31.4        \\
    10f          &24        &44.6        &184         &31.6        \\
                &          &            &            &            \\[-3mm]
F120:11a         &10        &125.3       &1.9         &8.1         \\
    11b          &28        &134.0       &209         &32.9        \\
    11c          &36        &134.6       &331         &38.3        \\
    11d          &60        &121.3       &463         &42.9        \\
    11e          &67        &109.1       &463         &42.9        \\
    11f          &76        &94.0        &463         &42.9        \\[-3mm]
  \end{tabular}
\end{center}
\tablenotetext{}{{\sc Note.}--- $M_{grid}$ is the total mass within
the computational grid, including the stellar mass $M_*$.}
\label{Tparameters}
\end{table}

In the following four figures the distributions of the gas density,
of the temperatures of amorphous carbon and silicate grains and
of the gas velocity are shown for selected cases (F60, G60, F30,
and F120) at six selected times.  The evolutionary age, total mass
within the computational grid, total luminosity (including
accretion luminosity), and mass of the central star which correspond
to each frame are given in Table \ref{Tparameters}.

\subsubsection{Case F60}

\begin{figure*}[htbp]
\epsscale{1.00}
\plotone{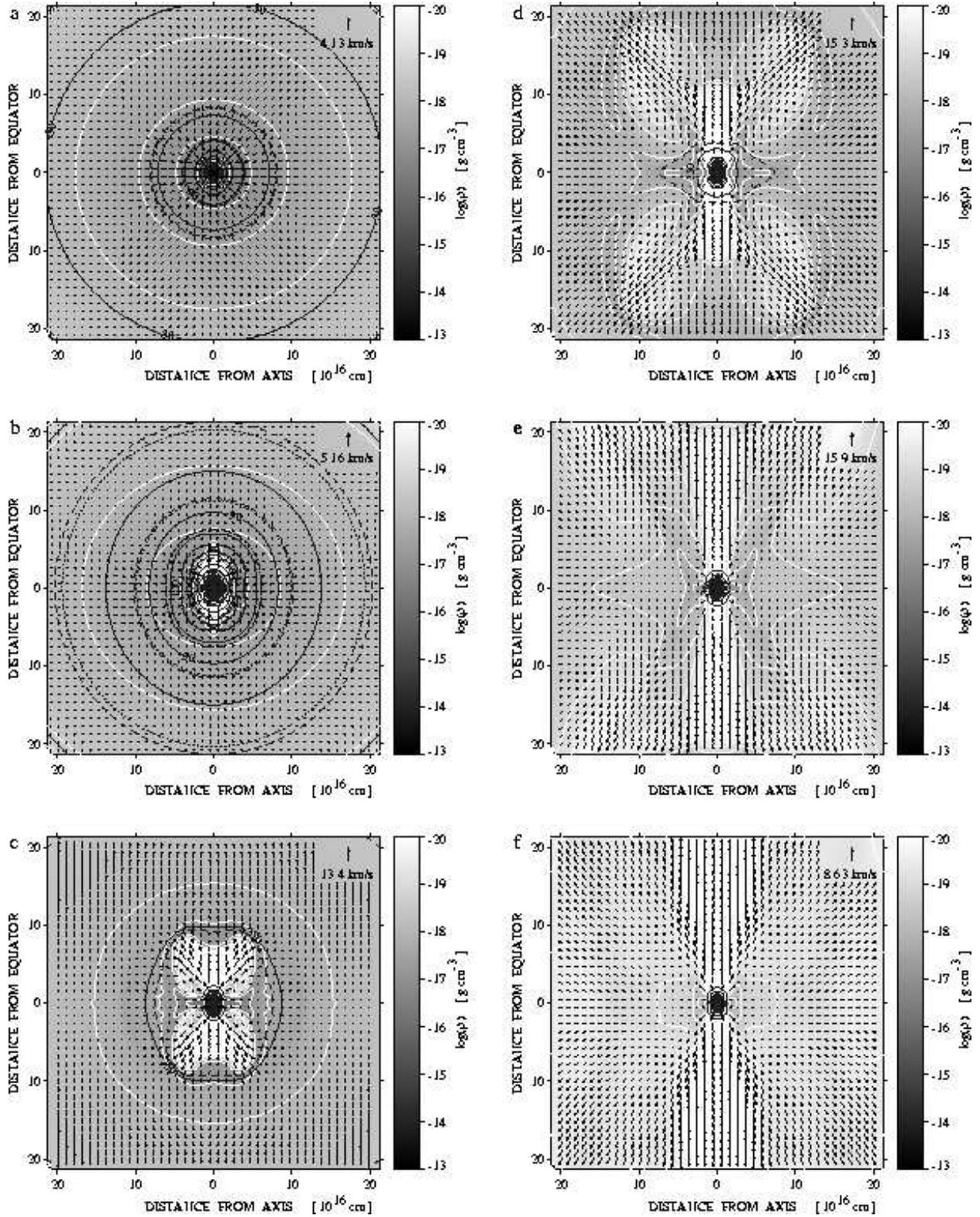}
\caption[]{Distribution of density ({\it grey-scale} and
{\it white contour lines}), velocity ({\it arrows}), temperature
of amorphous carbon grains ({\it solid black contour lines}), and
temperature of silicate grains ({\it dotted contour lines})
for case F60 (see Table \ref{Tinitial}) at evolutionary times
as indicated in Table \ref{Tparameters}. 
}
\label{fig-F60}
\end{figure*}

The evolution of the molecular clump considered in case F60 is
shown in Fig. \ref{fig-F60}.  Initially, the infall is nearly
spherically symmetric. Even after 10\,000 years (frame \ref{fig-F60}a)
at a time when the core mass has grown to 13.4~M\sol, there are
only minor differences in the
clump's structure along the equator compared to along the rotational
axis, evidenced by slight flattening of density contours and elongation
of temperature contours in the inner regions.  After 20\,000 years
(frame \ref{fig-F60}b) the non-sphericity has become more pronounced.
In the polar regions directly above and below the central object (a
28.4~M\sol\ star) the infall has been reversed by
radiative forces on the dust,  The low density outflowing region
is encased by shock fronts. The star continues to accrete
material through the equator.  After 25\,000 years (frame \ref{fig-F60}c)
the stellar mass has grown to 33.0~M\sol\ and the cavity
emptied by radiation has grown to $\pm 10^{17}$~cm in the polar
direction.  Its expansion velocity has also grown
($\ga 10$~km~s$^{-1}$).  Outside the cavity the infall of
clump material has also been reversed; only in the equatorial
plane is some infall possible.

For the last three frames of Fig. \ref{fig-F60} the central star
has reached the main sequence at 33.6~M\sol\ and no longer accretes.
Note that there is some indication of a flattened density structure
(see white contours in frame \ref{fig-F60}c), which persists for
the remainder of the evolution.  The highest expansion velocities
(and lowest densities) occur towards the poles.  Although 
$\sim 40$~M\sol\ of material was available within the computational
grid at 25\,000~years, radiative forces have effectively
prevented significant accretion after this time.  We stop the
computations at an evolutionary age of 45\,000~years
(frame \ref{fig-F60}f).

\begin{figure*}[htbp]
\epsscale{1.00}
\plotone{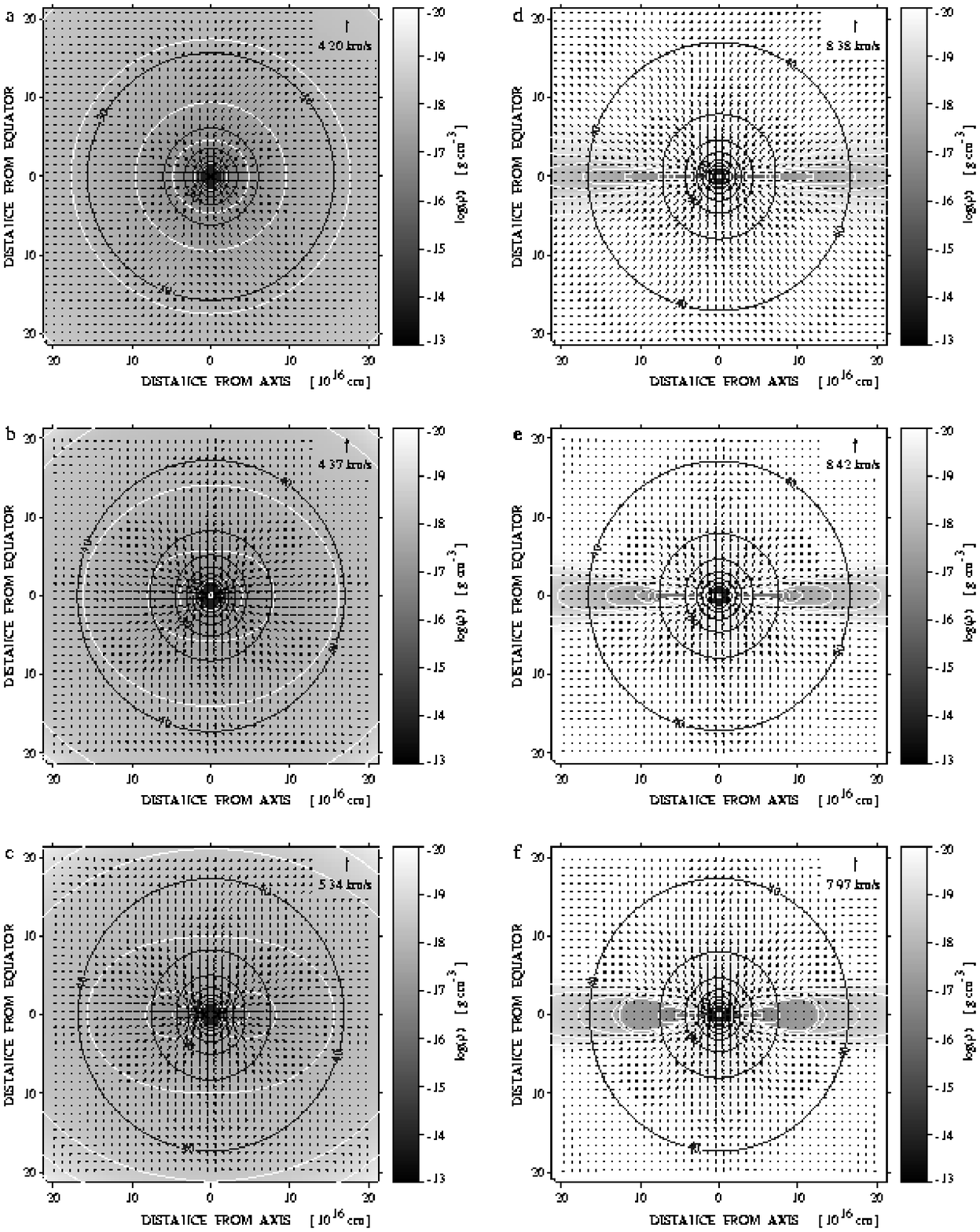}
\caption[]{Distribution of density, velocity, and grain
temperature for case G60 at evolutionary times
as indicated in Table \ref{Tparameters}.
Symbols and lines are as in Fig. \ref{fig-F60}.
}
\label{fig-G60}
\end{figure*}

\begin{figure*}[htbp]
\epsscale{1.00}
\plotone{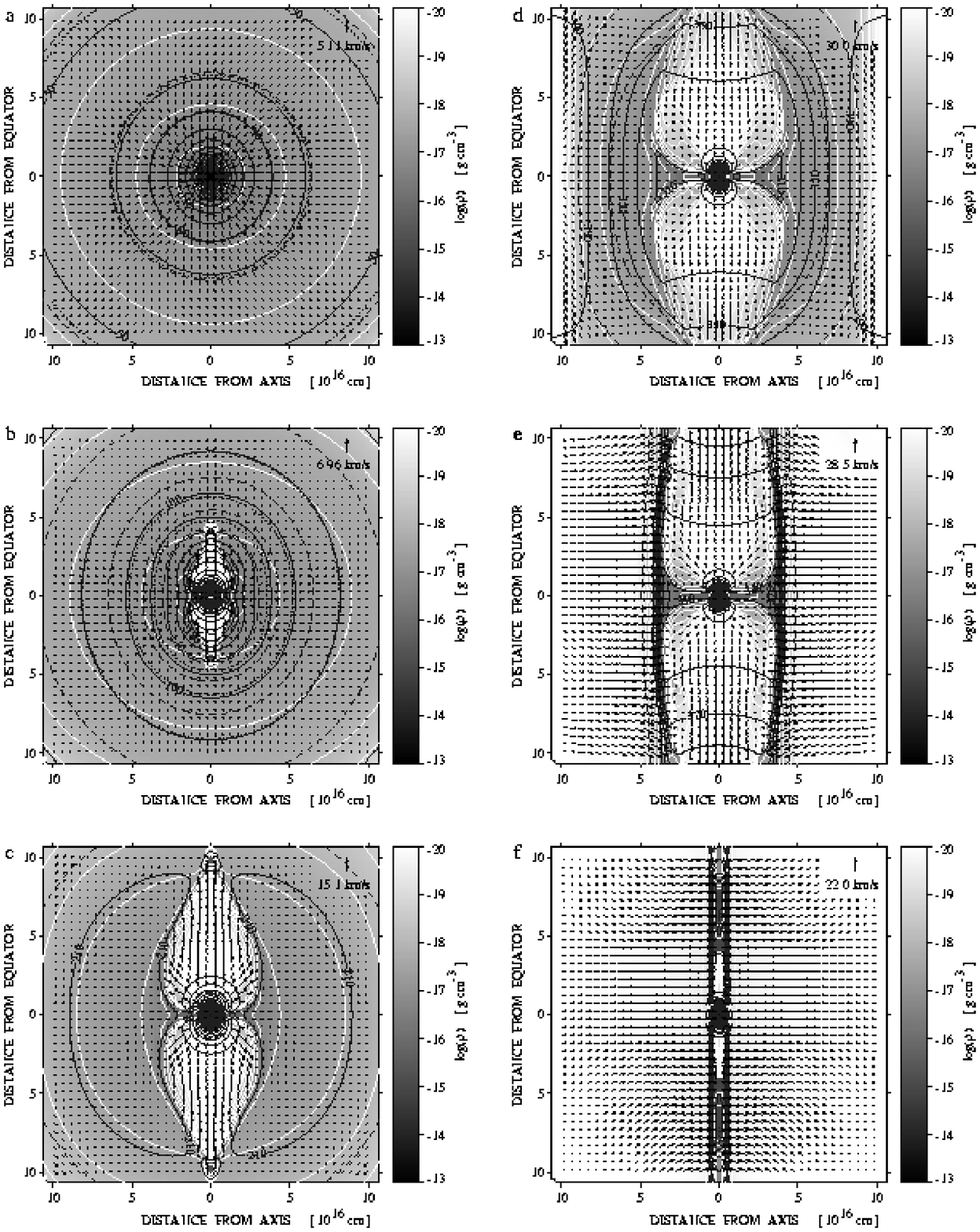}
\caption[]{Distribution of density, velocity, and grain
temperatures for case F30 at evolutionary times
as indicated in Table \ref{Tparameters}.
Symbols and lines are as in Fig. \ref{fig-F60}.
}
\label{fig-F30}
\end{figure*}

\begin{figure*}[htbp]
\epsscale{1.00}
\plotone{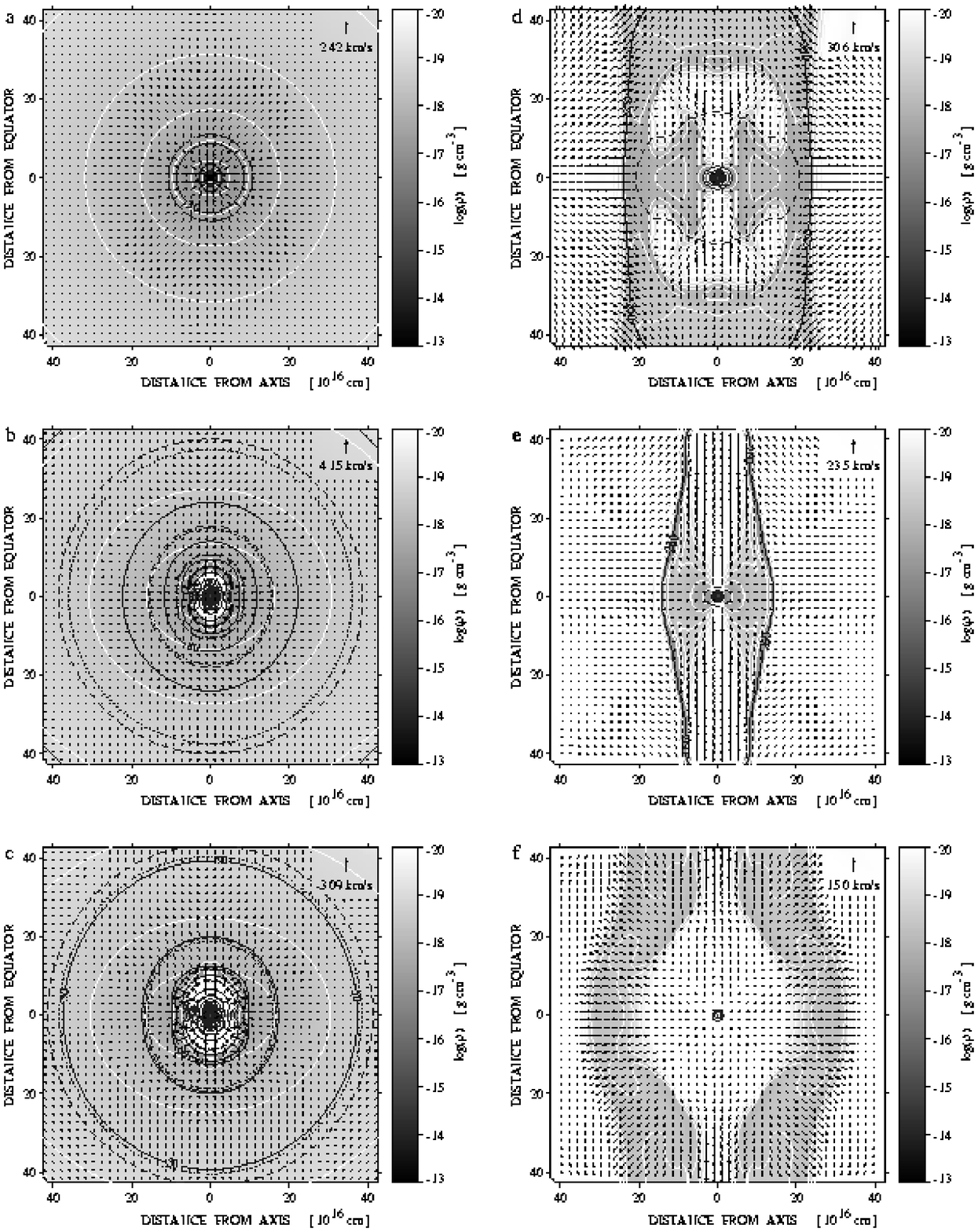}
\caption[]{Distribution of density, velocity, and grain
temperatures for case F120 at evolutionary times
as indicated in Table \ref{Tparameters}.
Symbols and lines are as in Fig. \ref{fig-F60}.
}
\label{fig-F120}
\end{figure*}

\subsubsection{Case G60}

Although the evolution of case G60 initially proceeds in a similar
fashion as case F60, notable differences are apparent after
25\,000~years (compare frame \ref{fig-G60}b to frame \ref{fig-F60}c).
The flow of material into the inner zone has effectively been
stopped at 15\,000~years, a time at which only 20.7~M\sol\ had been
accreted (see Fig. \ref{fig-mass60}). Contrary to case F60 there
is no indication of
the formation of a polar cavity, evacuated by radiative forces,
even at the most advanced evolutionary time considered, 110\,000~years
(frame \ref{fig-G60}f).  Instead, the material flows onto a thin, disk-like
structure, supported in the radial direction by both centrifugal and
radiative forces.

\subsubsection{Case F30}

As for case F60 the initial collapse of case F30 is nearly
spherically symmetric until an evacuated polar cavity is 
formed, encased in a system of expanding shock fronts.
Here, the infalling material collides with the radiatively
accelerated outflow.  In the condensed cylindrical shell
bounded by the shock fronts (see frame \ref{fig-F30}e) the
hard radiation from the central source is absorbed and
reemitted at longer wavelengths.  Because of this degrading of
the stellar radiation ``hardness'', the material outside the
cylindrical shell is able to flow radially inwards more or
less parallel to the equator for $|z| < 10^{17}$~cm.
Material continues to flow into the central zone via the
equatorial plane, and the central stellar mass ultimately
grows to 31.6~M\sol, more than originally present within the
computational grid.  The cylindrical shell depicted in frames
\ref{fig-F30}d and \ref{fig-F30}e is a short-lived phenomenon,
however.  Within a few thousand years it collapses into a 
long, narrow, filamentary structure (frame \ref{fig-F30}f)
containing about 13~M\sol.

\subsubsection{Case F120}

Again, as in cases F60 and F30 discussed above, the initial collapse
is spherically symmetric (frame \ref{fig-F120}a), followed by the
formation of a polar cavity evacuated by radiative forces (frame
\ref{fig-F120}b), after a significant amount of material has
accumulated within the central zone ($M_* = 32.9$~M\sol\ at
$t=28\,000$~yr.).  The central star continues to accrete an
additional 10~M\sol\ via an equatorial flow through a disk-like
structure over the next 30\,000 years albeit at an ever decreasing
rate.  This ``disk'' is short-lived, however.  At an
evolutionary age of 60\,000~years (frame \ref{fig-F120}d) the
accretion process has stopped and a cylindrical shell first forms,
contracts to a smaller cylindrical radius with a more focused
polar outflow (frame \ref{fig-F120}e), and then reexpands
(frame \ref{fig-F120}f). Whereas for earlier evolutionary times
the gas density was higher in the equatorial regions than in the
polar outflow regions, the final frame strongly resembles an
expanding ``cocoon'' shell, punctured and elongated by a polar
outflow.  No disk-like structure is visible.

\subsection{Appearance of Molecular Clumps}

\begin{figure}[tbhp]
\epsscale{1.00}
\plotone{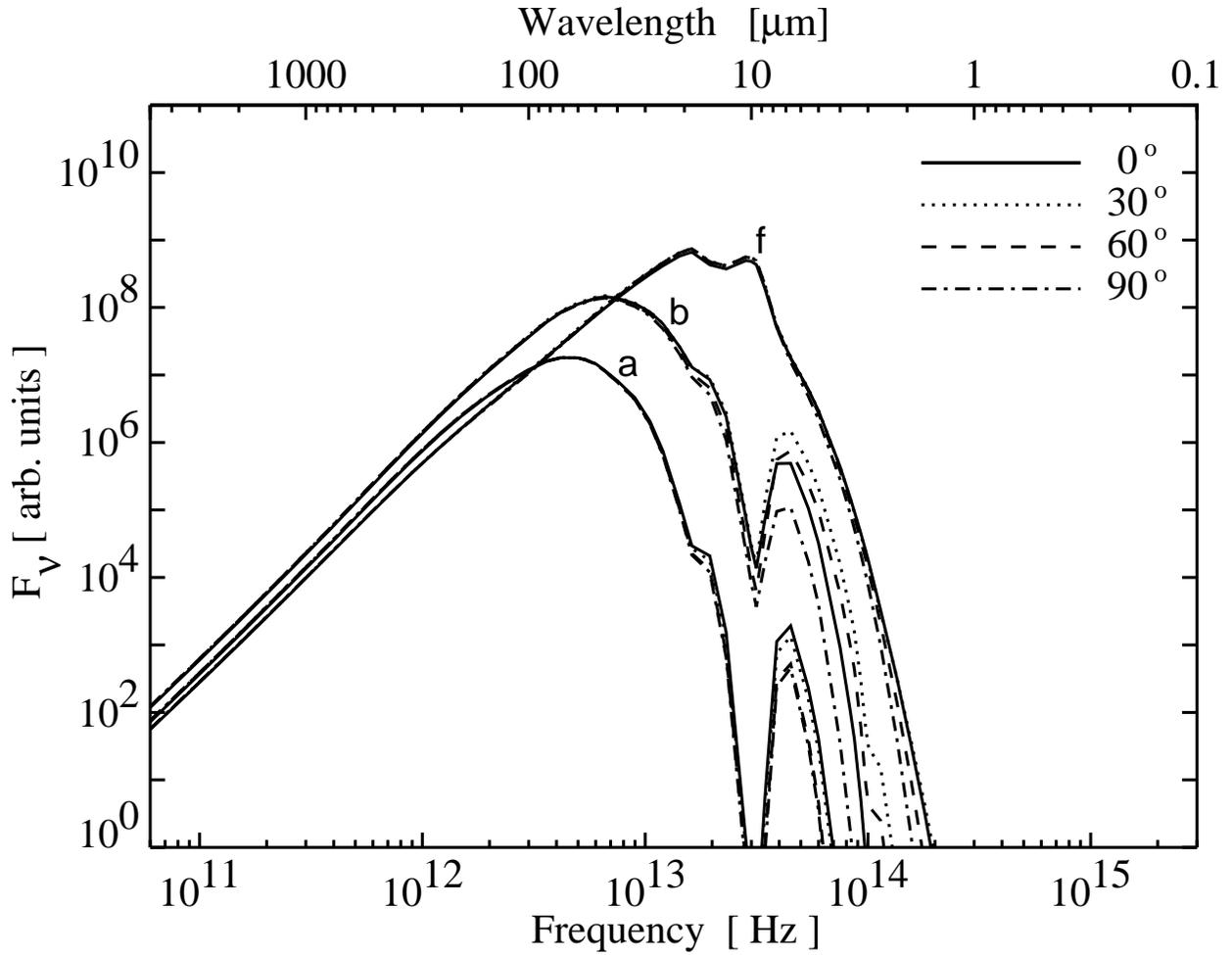}                                        
\caption[]{SEDs at selected evolutionary ages of sequence F60.
The four curves marked ``a'' correspond to the time shown in
frame \ref{fig-F60}a, ``b''
to frame \ref{fig-F60}b, and ``f'' to frame \ref{fig-F60}f.
The viewing angle $0^\circ$ is pole-on, $90^\circ$ is edge-on.
}
\label{fig-F60sp}
\end{figure}

\begin{figure}[tbhp]
\epsscale{1.00}
\plotone{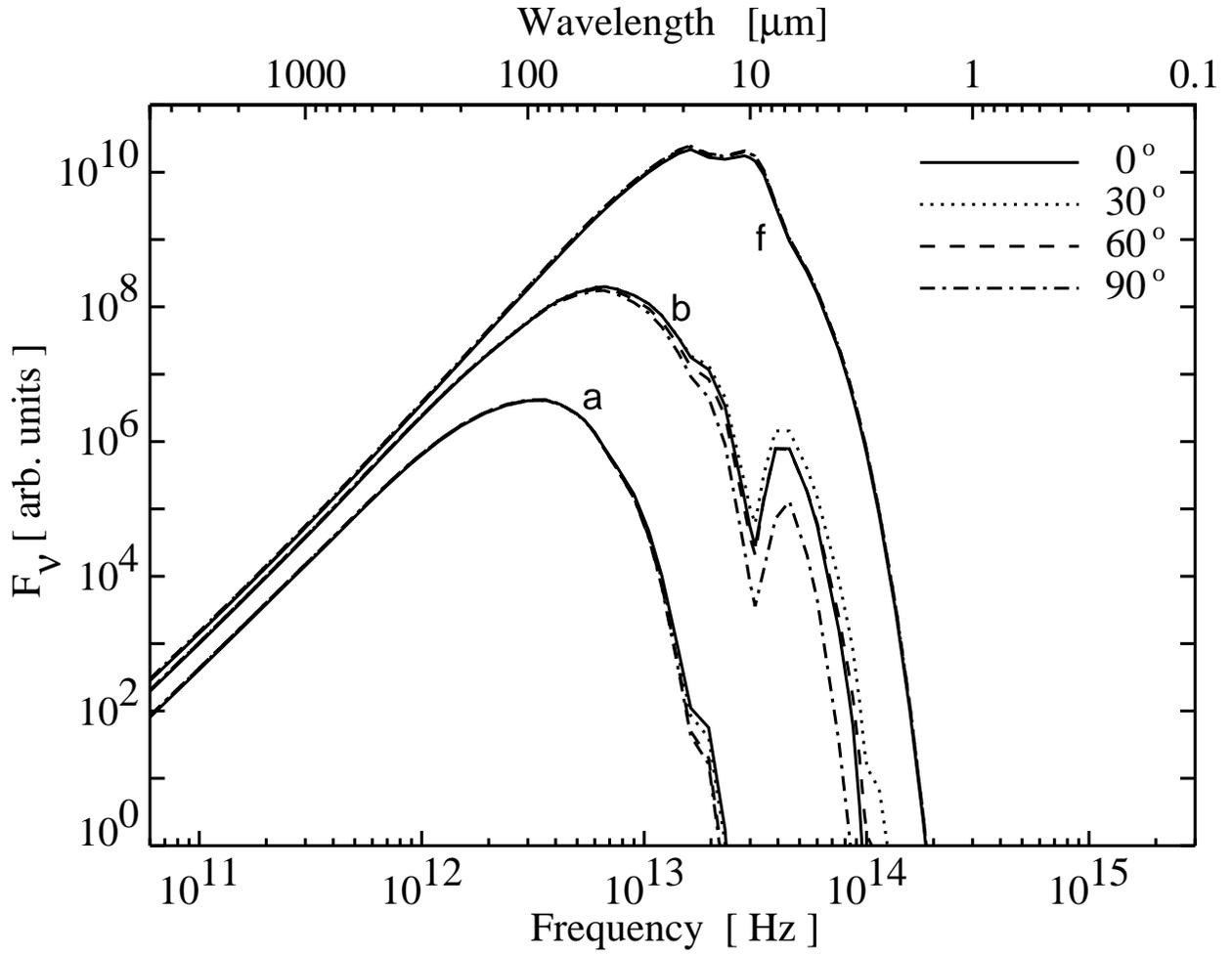}                                        
\caption[]{SEDs at selected evolutionary ages of sequence F120.
The four curves marked ``a'' correspond to the time shown in
frame \ref{fig-F120}a, ``b''
to frame \ref{fig-F120}b, and ``f'' to frame \ref{fig-F120}f.
}
\label{fig-F120sp}
\end{figure}

With the known density and equilibrium temperature distributions
of each dust component calculated for each time step, we can
perform ray-tracing radiation transfer calculations analogous to
those of YB, who --- by contrast --- used the single dust
temperature obtained in a grey radiation transfer code.
From these ray-tracing calculations we can extract both SEDs
(see Fig. \ref{fig-F60sp} and \ref{fig-F120sp})
and isophote maps at selected wavelengths for any given
evolutionary age.

Because, however, our spatial resolution in the innermost
regions is much worse than that of YB and furthermore, we
have made no attempt to model the emission from a hypothetical
disk within the central cell (see YB's ``central zone disk
model''), we will not be able to accurately model the emission
from dust warmer than about 400~K.  Even in the rather late
formation stages considered here --- the central stars have
evolved to main sequence O-stars, very little near infrared
and no optical/ultraviolet radiation escapes the remnant
molecular cloud.  It is reasonable to assume that a
non-homogeneous distribution of material within the central
zone (i.e. clumpiness or a flattened disk) would have allowed
at least some of the hard photons from the central source to
escape.  Thus, the examples given here can only be illustrative
of the basic method rather than an accurate model of the expected
near infrared to ultraviolet flux.  Unfortunately, our ignorance
of the goings-on within the central zone preclude a more
definitive treatment.

\section{Discussion and Conclusions}\label{conclusions}

Our improved frequency dependent radiation hydrodynamics code is
able to track the infall of material within a molecular clump
against radiative forces.  We find that the ``flashlight effect''
first discussed by Yorke \& Bodenheimer (1999), i.e. the
non-isotropic distribution of radiative flux that occurs when
a circumstellar disk forms, is strongly compounded by the
frequency dependent radiation transfer.
The shortest wavelength radiation (which is also the most effective 
for radiative acceleration) is most strongly concentrated towards
the polar directions, whereas the longer wavelength radiation (less
effective radiative acceleration) is more or less isotropic.
We conclude that
massive stars can in principle be formed via accretion through
a disk, in a manner analogous to the formation of lower mass
stars.  A powerful radiation-driven outflow in the polar
directions and a ``puffed-up'' (thick) disk result from the high
luminosity of the central source.

We have developed a simplified model for following the evolution
of accreting (proto-) stars, using existing tracks for non-accreting
stars.  With this model we have shown that in the case of massive
star formation the
energy released within the accretion shock front, the ``accretion
luminosity'', is {\sl not} the dominant source of luminosity after
a few thousand years of evolution.

The accretion rate onto the central source is time-dependent.  It
rises sharply after one free-fall time to a maximum value and falls
off gradually (in the frequency-dependent cases).  This is in
contrast to the expectations of Meynet \& Maeder (2000) and Behrend
\& Maeder (2001), who have assumed mass accretion rates $\dot M_*$
that increase monotonically in time up to a maximum value.

We have also shown that the
concept of ``birthline'', the equilibrium position of fully 
convective, deuterium-burning stars in the HR diagram with
cosmic deuterium abundance, is --- strictly speaking ---
unattainable for stars more massive
than 1~M\sol.  Beginning with a protostar of a fraction of a solar
mass and building up via accretion to 1~M\sol\ and higher masses, it
either accretes too rapidly (shifting the
HR position to smaller radii) or it accretes too slowly (significant
amounts of previously accreted deuterium are consumed).  For masses
$M \la 10$~M\sol, however, the contribution of the accretion
luminosity may make the star {\sl appear} to lie on or above the
birthline.

In this investigation we have not addressed the issues of the
longevity of the circumstellar disk or the possible formation
of a dense stellar cluster within our central computational zone
rather than a single star.  However, even without the assumption
of ionizing radiation, we find that these disks are not long-lived
phenomena.  In the most massive cases the effects of radiative
acceleration eventually disperse the remnant disks.  Future
studies will have to address the issues of ionization and the
interactions of the
disk with powerful stellar winds.  The effects of nearby companions
in a dense stellar cluster will also have to be considered in
future work.

\acknowledgements This research has been supported by the Deutsche
Forschungsgemeinschaft (DFG) within the framework of the ``Physics of
Star Formation'' program under grant Yo 5/14-3 and by the National 
Aeronautics and Space Administration (NASA) under the auspices of
the ``Origins'' Program and grant NRA-99-01-ATP-065.
Portions of this research were conducted
at Jet Propulsion Laboratory, California Institute of Technology.
The calculations were performed at the John von Neumann
Institute for Computing (NIC) in J\"ulich, at the Leibniz Computing
Center (LRZ) in Munich, and at the Astronomical Institute in W\"urzburg.


\begin{thebibliography}{}
\bibitem[Adams \& Shu 1986]{adams:shu86}
 Adams, F.C., \& Shu,  F.H.\ 1986, \apj, 308, 836
\bibitem[Allen 1973]{allen73}
 Allen, C.W.\ 1973, Astrophysical Quantities, 3rd ed., London: Athlone Press
\bibitem[Andr\'e et al. 2000]{andre00}
 Andr\'e, P., Ward-Thompson, D., Barsony, M.\ 2000,
 in Protostars \& Planets IV, ed. V. Mannings, A.P. Boss \& S.S. Russell,
 Tucson: Univ. of Arizona Press, p. 59
\bibitem[Andr\'e et al. 1996]{andre96}
 Andr\'e, P., Ward-Thompson, D., \& Motte, F.\ 1996, \aap, 314, 625
\bibitem[Auer 1987]{auer87}
 Auer, L.\ 1987, in Numerical Radiative Transfer, ed. W. Kalkofen,
 Cambridge: Cambridge Univ. Press, p. 101
\bibitem[Behrend  \& Maeder 2001]{behrend:maeder01}
 Behrend, R., Maeder, A.\ 2001, A\&A, 373, 190
\bibitem[Berger \& Colella 1989]{berger:colella89}
 Berger,  M.J. \& Colella, P. 1989, J. Comp. Phys., 82, 64
\bibitem[Bonnell et al. 1998]{bonnell98}
 Bonnell, I.A., Bate, M.R., Zinnecker, H.\ 1998, MNRAS, 298, 93
\bibitem[B\"ucker \& Sauren 1996]{buecker:sauren96}
 B\"ucker, H.M. \& Sauren, M.\ 1996, Internal Report kfa-zam-ib-9605,
 Research Centre J\"ulich
\bibitem[Cabrit \& Bertout 1992]{cabrit92}
Cabrit, S., \& Bertout, C.\ 1992, A\&A, 261, 274
\bibitem[Cannon 1973a]{cannon73a}
 Cannon, C.J.\ 1973a, J. Quant. Spectrosc. Rad. Trans., 13, 627
\bibitem[Cannon 1973b]{cannon73b}
 Cannon, C.J.\ 1973b, \apj, 185, 621
\bibitem[Churchwell 1998]{churchwell00}
 Churchwell, E.\ 1998, in {\sl The Origin of Stars and Planetary Systems},
 eds. C. Lada \& N. Kylafis, NATO Science Series, 540 (Kluwer), p. 515
\bibitem[Crutcher et al. 1994]{crutcher94}
 Crutcher, R., Mouschovias, T.Ch., Troland, T., \& Ciolek, G.\ 1994,
 \apj, 427, 839
\bibitem[D'Antona \& Mazzitelli 1994]{dantona:mazzitelli94}
D'Antona, F., \& Mazzitelli, I.\ 1994, \apjs, 90, 467
\bibitem[Draine \& Lee 1984]{draine:lee84}
 Draine, B.T., Lee, H.M.\ 1984, \apj, 285, 89
\bibitem[Eiroa et al. 1994]
 Eiroa, C., Casali, M. M., Miranda, L. F., Ortiz, E.\ 1994,
 A\&A,  290, 599
\bibitem[Henning 2000]{henning00}
 Henning, T., Schreyer, K, Launhardt, R., Burkert, A.\ 2000,
 A\&A, 353, 211
\bibitem[Hollenbach et al. 2000]{hollenbach00}
 Hollenbach, D., Yorke, H.W., Johnstone, D.\ 2000,
 in Protostars \& Planets IV, ed. V. Mannings, A.P. Boss \& S.S. Russell,
 Tucson: Univ. of Arizona Press, p. 401
\bibitem[Iben 1965]{iben65}
 Iben, I., Jr.\ 1965, \apj, 141, 993
\bibitem[Kippenhahn \& Hofmeister 1977]{kippenhahn:hofmeister77}
Kippenhahn, R.; Meyer-Hofmeister, E.\ 1977, \aap, 54, 539
\bibitem[Kippenhahn \& Weigert 1980]{kippenhahn:weigert90}
 Kippenhahn, R., Weigert, A.\ 1980, Stellar Structure and Evolution,
 Heidelberg: Springer Verlag
\bibitem[Levermore \& Pomraning 1981]{levermore:pomraning81}
 Levermore, C., Pomraning, G.\ 1981, \apj, 248, 321
\bibitem[Lizano \& Shu 1989]{lizano89}
 Lizano, S., Shu, F.H.\ 1989, ApJ, 342, 834
measurements Martin-Pintado et al. (1994) do find indirect 
\bibitem[Martin-Pintado et al. 1994]{martin-pintado94}
 Martin-Pintado, J., Neri, R., Thum, C., Planesas, P., Bachiller, R.,
 1994 A\&A, 286, 890
\bibitem[Mathis, Rumpl, \& Nordsieck 1977]{mathis77}
 Mathis, J.S., Rumpl, W., Nordsieck, K.H.\ 1977,  \apj, 217, 425
\bibitem[Meynet \& Maeder 2000]{meynet:maeder00}
 Meynet, G., Maeder, A.\ 2000, A\&A, 361, 101
\bibitem[Motte et al. 1998]{motte98}
 Motte, F., Andr\'e, P., \& Neri, R.\ 1998, \aap, 336, 150
\bibitem[Mouschovias 1990]{mouschovias90}
 Mouschovias T.Ch. 1990, in Physical Processes in Fragmentation and
 Star Formation, ed. R. Capuzzo-Dolcetta, C. Chiosi, \& A. Di Fazio,
 Dordrecht: Kluwer, p. 117
\bibitem[Preibisch et al. 1993]{preibisch93}
 Preibisch, T., Ossenkopf, V., Yorke, H.W., Henning, T.\ 
 1993, \aap, 279, 577
\bibitem[Press et al. 1992]{press92}
 Press, W.H., Teukolsky, S.A., Vetterling, W.T., Flannery, B.P.\ 1992,
 Numerical Recipes in C, 2nd ed., Cambridge: Cambridge Univ. Press
\bibitem[Richling \& Yorke 1998]{richling:yorke98}
 Richling, S., Yorke, H.W.\ 1998, \aap, 340, 508 
\bibitem[Richer et al. 2000]{richter00}
 Richer, J. S., Shepherd, D. S., Cabrit, S., Bachiller, R.,
 Churchwell, E.\ 2000,
 in Protostars \& Planets IV, ed. V. Mannings, A.P. Boss \& S.S. Russell,
 Tucson: Univ. of Arizona Press, p. 867
\bibitem[Rozyczka et al. 1994]{rozyczka94}
 R\'o\.zyczka, M., Bodenheimer, P., \& Bell, K.R.\ 1994, \apj, 
 423, 736
\bibitem[Scharmer 1981]{shcarmer81}
 Scharmer, G.B.\ 1981, \apj, 249, 720
\bibitem[Shakura \& Sunyaev 1973]{shakura:sunyaev73}
 Shakura, N.I., Sunyaev, R.A. 1973, \aap, 24, 337
\bibitem[Shepherd \& Churchwell 1996]{shepard96}
 Shepherd, D. S., Churchwell, E.\ 1996, \apj, 472, 225
\bibitem[Shepherd et al. 2000]{shepard00}
 Shepherd, D. S., Yu, K. C., Bally, J., Testi, L.\ 2000,
 \apj, 535, 833
\bibitem[Sleijpen \& Fokkema 1993]{sleijpen:fokkema93}
 Sleijpen, G.L.G., Fokkema, D.R.\ 1993, Electr. Transactions on Num.
 Analysis, 1, 11 \hfil\break (etna@mcs.kent.edu)
\bibitem[Sonnhalter et al. 1995]{sonnhalter95}
 Sonnhalter, C., Preibisch, T., Yorke, H.W.\ 1995, \aap, 299, 545
\bibitem[Stahler et al. 2000]{stahler00}
 Stahler, S.W., Palla, F., Ho, P.T.P.\ 2000,
 in Protostars \& Planets IV, ed. V. Mannings, A.P. Boss \& S.S. Russell,
 Tucson: Univ. of Arizona Press, p. 327
\bibitem[Tomisaka et al. 1990]{tomisaka90}
 Tomisaka,  K., Ikeuchi,  S., \& Nakamura,  T.\ 1990, \apj, 362, 202
\bibitem[Williams et al. 2000]{williams00}
 Williams, J.P., Blitz, L., McKee, C.F.\ 2000,
 in Protostars \& Planets IV, ed. V. Mannings, A.P. Boss \& S.S. Russell,
 Tucson: Univ. of Arizona Press, p. 97
\bibitem[Yorke \& Bodenheimer 1999]{yorke:bodenheimer99}
 Yorke, H.W., Bodenheimer, P.\ [YB] 1999, \apj, 525, 330
\bibitem[Yorke et al. 1995]{yorke95}
 Yorke, H.W., Bodenheimer, P., Laughlin, G.\ 1995, \apj, 443, 199
\bibitem[Yorke \& Kaisig 1995]{yorke:kaisig95}
 Yorke, H.W., Kaisig, M.\ 1995, Comp.\ Phys.\ Comm., 89, 29
\bibitem[Yorke \& Henning 1994]{yorke:henning94}
 Yorke, H.W., Henning, T.\ 1994, in Molecules in the Stellar Environment,
 IAU Coll. No. 146, ed. U.G. J{\o}rgensen, Berlin: Springer, p. 186
\end{thebibliography}
\end{document}